\documentclass[conference]{IEEEtran}
\IEEEoverridecommandlockouts

\usepackage{cite}
\usepackage{amsmath,amssymb,amsfonts}
\usepackage{algorithmic}
\usepackage{graphicx}
\usepackage{textcomp}
\usepackage{xcolor}
\usepackage{hyperref}
\usepackage[font=normalsize,labelfont=bf,textfont=normal,justification=centering]{caption}
\usepackage{tikz}
\usetikzlibrary{shapes.geometric, arrows.meta, positioning, calc, decorations.pathreplacing, backgrounds, fit}
\usepackage{bbm}
\def\BibTeX{{\rm B\kern-.05em{\sc i\kern-.025em b}\kern-.08em
    T\kern-.1667em\lower.7ex\hbox{E}\kern-.125emX}}
\usepackage{subcaption}
\usepackage[pass]{geometry}
\begin{document}

\title{STCTS: Generative Semantic Compression for Ultra-Low Bitrate Speech via Explicit Text-Prosody-Timbre Decomposition}

\author{
\IEEEauthorblockN{Siyu Wang}
\IEEEauthorblockA{Fudan University\\
23300240006@m.fudan.edu.cn}
\and
\IEEEauthorblockN{Haitao Li}
\IEEEauthorblockA{Tsinghua University\\
liht22@mails.tsinghua.edu.cn}
\and
\IEEEauthorblockN{Donglai Zhu\IEEEauthorrefmark{1}}
\IEEEauthorblockA{Fudan University\\
zhudl@fudan.edu.cn}
}

\maketitle

\begingroup\renewcommand\thefootnote{\IEEEauthorrefmark{1}}
\footnotetext{Corresponding author. Address: Room D6009, No. 2 Interdisciplinary Building, Jiangwan Campus, Fudan University, Shanghai 200438, China. Tel: +86-13916747999. Email: zhudl@fudan.edu.cn}
\endgroup

\begin{abstract}
\setlength{\baselineskip}{1.2\baselineskip}
Voice communication in bandwidth-constrained environments—maritime, satellite, and tactical networks—remains prohibitively expensive. Traditional codecs struggle below 1 kbps, while existing semantic approaches (STT-TTS) sacrifice prosody and speaker identity. We present \textbf{STCTS}, a generative semantic compression framework enabling natural voice communication at \textbf{$\sim$80 bps}. STCTS explicitly decomposes speech into linguistic content, prosodic expression, and speaker timbre, applying tailored compression: context-aware text encoding ($\sim$70 bps), sparse prosody transmission via TTS interpolation ($<$14 bps at 0.1--1 Hz), and amortized speaker embedding.

Evaluations on LibriSpeech demonstrate a \textbf{$75\times$ bitrate reduction} versus Opus (6 kbps) and \textbf{$12\times$} versus EnCodec (1 kbps), while maintaining perceptual quality (NISQA MOS $>$ 4.26), graceful degradation under packet loss and noise resilience. We also discover a \textbf{bimodal quality distribution} with prosody sampling rate: sparse and dense updates both achieve high quality, while mid-range rates degrade due to perceptual discontinuities—guiding optimal configuration design. Beyond efficiency, our modular architecture supports privacy-preserving encryption, human-interpretable transmission, and flexible deployment on edge devices, offering a robust solution for ultra-low bandwidth scenarios.
\end{abstract}

\begin{IEEEkeywords}
\setlength{\baselineskip}{1.2\baselineskip}
Ultra-low bitrate communication, semantic compression, speech-to-text, text-to-speech, prosody encoding, voice cloning, bandwidth-constrained networks, neural speech synthesis
\end{IEEEkeywords}

\section{Introduction}

\begin{figure}[t]
\centering
\includegraphics[width=\columnwidth]{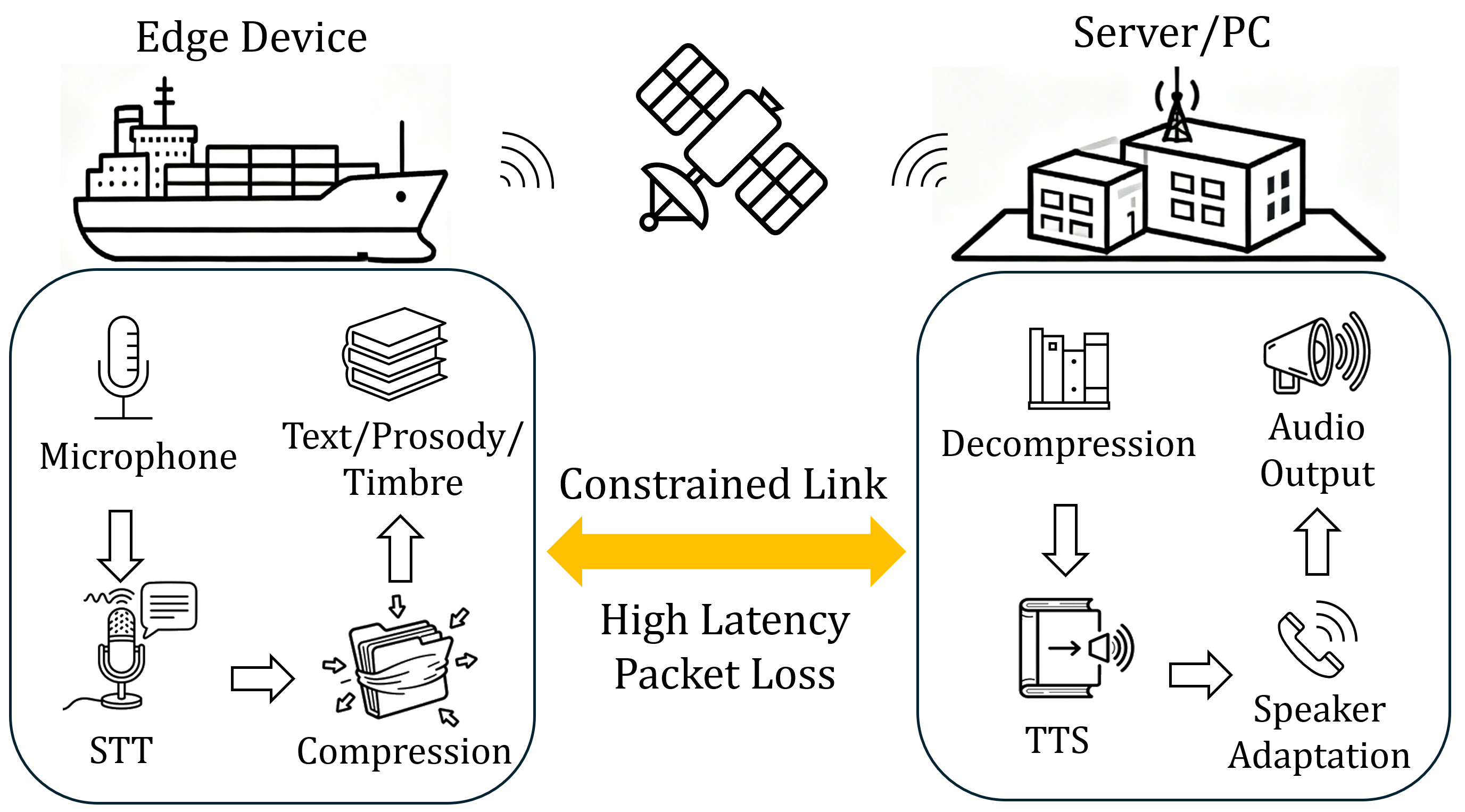}
\caption{Operational scenario. Edge devices (e.g., on maritime vessels) transmit decomposed speech components (text, prosody, timbre) over constrained satellite links. The receiver reconstructs natural speech from these semantic streams, overcoming high latency and packet loss.}
\label{fig:scenario}
\end{figure}

Voice communication remains a fundamental human need, yet in many regions and circumstances worldwide, network bandwidth is severely constrained and prohibitively expensive. Consider maritime workers aboard cargo ships and fishing vessels: they rely on satellite communication systems where bandwidth costs can reach \$5--\$15 per megabyte, making a single 10-minute voice call at standard telephony bitrates (e.g. 20 -- 30 kbps) cost approximately \$20---a substantial burden for workers often earning modest wages. Consequently, crew members are often restricted to brief, infrequent calls home, exacerbating the isolation inherent to months-long voyages. Similar bandwidth constraints afflict satellite and aeronautical communication systems (e.g., in-flight connectivity, drone control links), wireless networks in remote and rural regions (e.g., Sub-Saharan Africa, Pacific islands, Himalayan communities), tactical military communication systems operating under contested spectrum conditions, and emerging large-scale real-time voice social platforms seeking to serve millions of concurrent users with minimal infrastructure costs. In all these scenarios, the fundamental question remains: \textit{how can we enable natural, affordable voice communication with minimal bandwidth consumption?}

To achieve natural, expressive voice communication at ultra-low bitrates, we draw insights from three distinct research domains, each offering valuable principles while facing fundamental limitations:

\textbf{1. Existing Speech Coding and Semantic Compression Techniques.} Traditional speech codecs compress acoustic waveforms through parametric modeling (e.g., Opus at 6--40 kbps \cite{opus}) or neural encoding (e.g., EnCodec at 1--24 kbps \cite{encodec}). While achieving significant compression compared to uncompressed audio (64--128 kbps), they remain fundamentally limited by waveform-level fidelity preservation, preventing operation below $\sim$1 kbps. Recent semantic compression methods \cite{collette2024,semanticodec} (e.g., Vevo at $\sim$650 bps) demonstrate that encoding speech at higher abstraction levels—representing \textit{what is said} and \textit{how it is said} through discrete tokens rather than raw acoustics—enables order-of-magnitude compression gains while maintaining perceptual quality through generative reconstruction. However, token-based representations lack interpretability and modularity: transmitted content is opaque to inspection, and upgrading individual components (recognition or synthesis models) requires end-to-end retraining of the entire system.

\textbf{2. STT-TTS Communication Systems.} STT-TTS pipelines \cite{urazayev2024voice} in IoT and tactical communication scenarios where bandwidth is severely constrained achieve ultra-low bitrates by converting speech to text, transmitting the compressed text ($\sim$70 bps), and resynthesizing speech at the receiver. The use of explicit text representation provides significant advantages: transmitted content is human-readable and debuggable, STT and TTS components can be independently upgraded as technology advances, and the architecture naturally enables secondary applications such as real-time transcription and multilingual translation. However, transmitting only linguistic content sacrifices two essential dimensions of human communication—\textit{prosodic expressiveness} (intonation, emphasis, emotion) and \textit{speaker identity} (voice timbre, characteristics). In maritime communication, for instance, crew members calling home expect their families to recognize their voice and perceive their emotional state; a generic synthesized voice transmitting only words feels impersonal and detached.

\textbf{3. Speech Disentanglement for Representation Learning.} Speech disentanglement approaches \cite{speechtriplenet,speechsplit} decompose speech into orthogonal representations corresponding to content, prosody, and timbre through end-to-end learned encoders (hereafter, ``content'' and ``text'' are used interchangeably in this paper). This factorization has proven effective for voice conversion and controllable synthesis tasks, revealing a crucial insight: speech components exhibit vastly different temporal dynamics—linguistic content changes rapidly (2--3 words/sec), prosody varies smoothly over multi-second spans, and speaker identity remains constant across conversations. This understanding of component-specific temporal structure is fundamental to designing effective compression strategies. However, these methods target representation learning rather than communication—they transmit continuous frame-level latent vectors at 50--100 Hz (requiring hundreds to thousands of bps) and produce learned representations that lack the interpretability and modularity needed for practical deployment in bandwidth-constrained scenarios.

\begin{figure*}[t]
\centering
\begin{tikzpicture}[
    node distance=0.7cm and 1.2cm,
    every node/.style={font=\small},
    block/.style={rectangle, draw=black!60, thick, fill=blue!8, text width=2.2cm, align=center, minimum height=0.75cm, rounded corners=2pt},
    compress/.style={rectangle, draw=black!60, thick, fill=green!12, text width=2.2cm, align=center, minimum height=0.7cm, rounded corners=2pt},
    channel/.style={rectangle, draw=black!70, thick, fill=orange!8, text width=2.3cm, align=center, minimum height=1cm},
    decompress/.style={rectangle, draw=black!60, thick, fill=purple!8, text width=2.2cm, align=center, minimum height=0.7cm, rounded corners=2pt},
    database/.style={cylinder, draw=black!60, thick, fill=yellow!12, shape border rotate=90, aspect=0.3, text width=1.6cm, align=center, minimum height=0.9cm},
    arrow/.style={-Stealth, line width=1pt},
    sparrow/.style={-Stealth, line width=1pt, dashed},
    oncearrow/.style={-Stealth, line width=1pt, dotted, line width=1.2pt},
    label/.style={font=\scriptsize\itshape, text=black!60},
    senderbox/.style={draw=blue!60, line width=1.5pt, rounded corners=4pt, dashed, inner sep=10pt},
    receiverbox/.style={draw=purple!60, line width=1.5pt, rounded corners=4pt, dashed, inner sep=10pt}
]

\node[block] (audio) {Audio Input\\{\scriptsize(16 kHz)}};
\node[block, below=0.5cm of audio] (vad) {VAD};

\node[block, below left=0.9cm and 2cm of vad] (stt) {STT\\Model};
\node[block, below=0.9cm of vad] (prosody) {Prosody\\Extraction};
\node[block, below right=0.9cm and 2cm of vad] (embedding) {Speaker\\Embedding};

\node[compress, below=0.6cm of stt] (textcomp) {Context-aware\\Compression};
\node[compress, below=0.6cm of prosody] (prosodycomp) {Quantization +\\Delta Encoding};
\node[compress, below=0.6cm of embedding] (timbrecomp) {Universal\\Compression};

\coordinate (chan_y) at ($(textcomp.south)+(0,-2.0cm)$);
\node[channel, anchor=north] (textchan) at (textcomp.south |- chan_y) {\textbf{HIGH} Priority\\Continuous\\{\scriptsize$\sim$70 bps}};
\node[channel, anchor=north] (prosodychan) at (prosodycomp.south |- chan_y) {\textbf{MED/LOW}\\Every 0.1--1s\\{\scriptsize 0.7--14 bps}};
\node[channel, anchor=north] (timbrechan) at (timbrecomp.south |- chan_y) {\textbf{HIGH} Priority\\Once/speaker\\{\scriptsize(amortized)}};

\coordinate (decomp_y) at ($(textchan.south)+(0,-2.5cm)$);
\node[decompress, anchor=north] (textdecomp) at (textchan.south |- decomp_y) {Decompress};
\node[decompress, anchor=north] (prosodydecomp) at (prosodychan.south |- decomp_y) {Decompress +\\Interpolation};
\node[decompress, anchor=north] (timbredecomp) at (timbrechan.south |- decomp_y) {Decompress};

\node[database, right=0.4cm of timbredecomp] (cache) {Timbre\\Profiles\\{\scriptsize(cached)}};

\node[block, below=1.1cm of prosodydecomp, text width=3cm] (tts) {TTS Synthesis};

\node[block, below=0.5cm of tts] (output) {Reconstructed Audio};

\draw[arrow] (audio) -- (vad);
\draw[arrow] (vad.south) -- ++(0,-0.3) -| (stt.north);
\draw[arrow] (vad) -- (prosody);
\draw[arrow] (vad.south) -- ++(0,-0.3) -| (embedding.north);

\draw[arrow] (stt) -- (textcomp);
\draw[arrow] (textcomp) -- (textchan) node[midway, right, label] {TEXT};
\draw[arrow] (textchan) -- (textdecomp) node[pos=0.3, right, font=\scriptsize\itshape, text=red!70] {Retrans. on fail};
\draw[arrow] (textdecomp.south) -- ++(0,-0.75) -| (tts.150);

\draw[sparrow] (prosody) -- (prosodycomp);
\draw[sparrow] (prosodycomp) -- (prosodychan) node[midway, right, label] {PROSODY};
\draw[sparrow] (prosodychan) -- (prosodydecomp) node[pos=0.3, right, font=\scriptsize\itshape, text=red!70] {Interp. if lost};
\draw[sparrow] (prosodydecomp) -- (tts);

\draw[oncearrow] (embedding) -- (timbrecomp);
\draw[oncearrow] (timbrecomp) -- (timbrechan) node[midway, right, label] {TIMBRE};
\draw[oncearrow] (timbrechan) -- (timbredecomp);
\draw[oncearrow] (timbredecomp) -- (cache);
\draw[oncearrow] (cache.south) -- ++(0,-0.3) -| (tts.30);

\draw[arrow] (tts) -- (output);

\node[above left=0cm and 3cm of audio, font=\Large\bfseries] (senderlabel) {Sender};
\node[below left=-1.5cm and 0cm of textdecomp, font=\Large\bfseries] (receiverlabel) {Receiver};

\coordinate (sender_bottom) at (textcomp.south);
\coordinate (receiver_top) at (textdecomp.north);
\coordinate (vertical_mid) at ($(sender_bottom)!0.5!(receiver_top)$);

\node[left=5cm of vertical_mid, rotate=90, font=\Large\bfseries, anchor=center] (channellabel) {Low-Bandwidth Channel};

\node[right=5cm of prosodychan, font=\small, text=orange!70, align=center] (webrtc) {WebRTC\\Data\\Channel};

\begin{scope}[on background layer]
    \node[senderbox, fit={(senderlabel) (audio) (vad) (stt) (prosody) (embedding) (textcomp) (prosodycomp) (timbrecomp)}] {};
    \node[receiverbox, fit={(receiverlabel) (textdecomp) (prosodydecomp) (timbredecomp) (cache) (tts) (output)}] {};
    \node[draw=orange!50, line width=1.5pt, rounded corners=4pt, fill=orange!2, fit={(textchan) (prosodychan) (timbrechan)}, inner sep=8pt] {};
\end{scope}

\end{tikzpicture}
\caption{STCTS system architecture. The sender decomposes speech into three orthogonal components—text (continuous, $\sim$70 bps), prosody (sparse keyframes, 0.7--14 bps), and timbre (one-time transmission, amortized)—each compressed with tailored strategies. These are transmitted via WebRTC data channels with prioritized delivery. The receiver reconstructs natural speech via TTS conditioning on all three components, with timbre profiles cached locally for recurring speakers.}
\label{fig:system_overview}
\end{figure*}
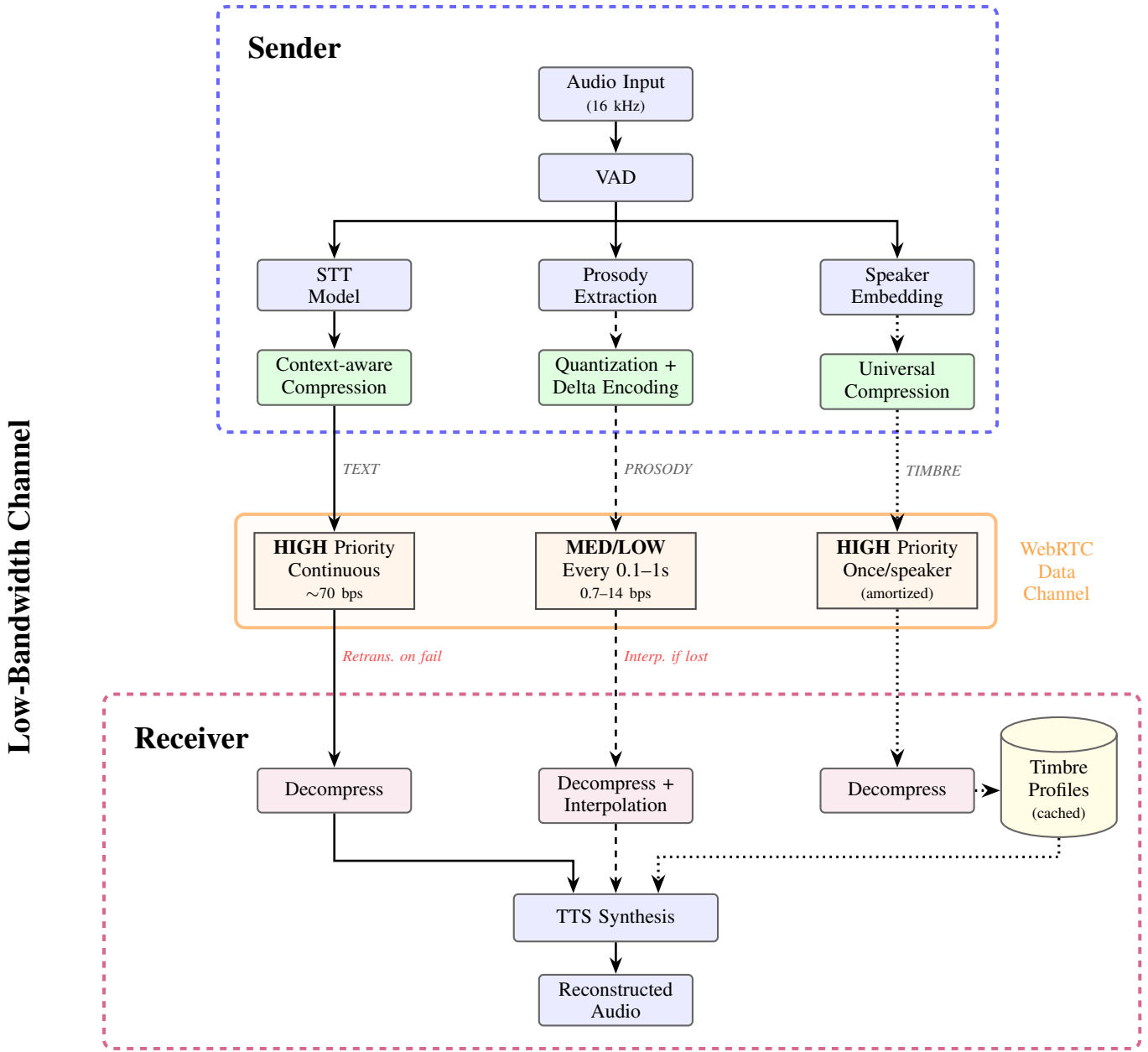

\textbf{Our Approach: STCTS.} Building on these insights, we present \textbf{STCTS} (\textbf{S}peech-\textbf{t}o-\textbf{T}ext, \textbf{C}ompression and \textbf{T}ext-\textbf{t}o-\textbf{S}peech), a framework achieving natural voice communication at \textbf{$\sim$80 bps} via explicit semantic decomposition. Our central insight is that \textit{speech can be reconstructed from high-level representations—content, prosody, and timbre—without preserving waveform fidelity.} This abstraction enables component-specific compression strategies to minimize bandwidth usage: \textbf{Linguistic content} is compressed via context-aware encoding ($\sim$70 bps); \textbf{Prosodic expression}, varying smoothly, allows sparse transmission (0.1--1 Hz) with TTS interpolation ($<$14 bps) and delta-encoded quantization; and \textbf{Speaker identity} requires only one-time amortized transmission per speaker. This explicit decomposition achieves a bitrate comparable to Morse code transmissions while conveying full conversational speech with near-natural quality and speaker fidelity—capabilities that prior semantic codecs (lacking interpretability), STT-TTS systems (lacking expressiveness), and disentanglement methods (requiring frame-level transmission) cannot simultaneously deliver. The architecture (Figure~\ref{fig:system_overview}) extracts these components at the sender and reconstructs them at the receiver using a conditioned TTS model. 

\textbf{Key Design Distinctions.} Our approach differs fundamentally from prior semantic compression work in two aspects. First, unlike token-based methods (e.g., Vevo \cite{collette2024}) that encode speech into discrete acoustic tokens, we use explicit text as the semantic representation. This design choice provides interpretability (transmitted content is human-readable), modularity (STT and TTS components can be independently upgraded), and enables secondary applications (real-time transcription, conversation logging, multilingual translation). Second, we transmit prosody features at extremely low rates (0.1--1 Hz, corresponding to updates every 1--10 seconds) by exploiting TTS models' ability to interpolate smooth prosody contours between sparse keyframes. Through systematic analysis of prosody sampling rates (Section~\ref{sec:prosody_analysis}), we identify a bimodal quality distribution with optimal operating points at sparse rates, enabling near-zero prosody bitrate ($<$14 bps) without sacrificing naturalness.

We conduct comprehensive experiments on the LibriSpeech corpus \cite{librispeech}, evaluating three quality modes (minimal, balanced, high-quality) against baseline codecs (Opus, EnCodec) and the Vevo semantic compression framework. Our contributions are as follows:

\begin{itemize}
\item We achieve sustained bitrates of 71.6--79.6 bps (excluding amortized speaker embeddings), representing a $75\times$ reduction compared to Opus (6 kbps) and an $8\times$ reduction compared to EnCodec (1 kbps), while maintaining perceptual quality (NISQA MOS $\sim$4.26) comparable to Vevo ($\sim$650 bps, MOS 4.21).

\item We demonstrate that prosody can be transmitted at extremely sparse rates (0.1--1 Hz, $<$14 bps) by leveraging TTS interpolation, and identify a bimodal quality distribution where mid-frequency prosody updates (1--5 Hz) perform worse than both sparse and dense regimes due to perceptually salient discontinuities.

\item We show that semantic compression exhibits inherent temporal desynchronization (reflected in low STOI scores $\sim$0.15) due to independent STT and TTS timing, yet maintains high intelligibility (WER $\sim$0.23) and naturalness (NISQA $>$4.2).

\item We achieve graceful degradation under channel noise (0.1--10\% bit error rate), maintaining NISQA MOS $>$4.2 even at 10\% BER through prioritized transmission and prosody interpolation, demonstrating robustness for real-world deployment in degraded communication environments.

\item We demonstrate the computational feasibility of our approach on consumer hardware, achieving a Real-Time Factor (RTF) of $\sim$0.4 for the full pipeline (STT / factorization, compression, decompression, TTS / reconstruction). This confirms that the system can operate comfortably in real-time on a single consumer-grade GPU, validating the practical viability of semantic compression for live communication.

\item We provide an open-source implementation with configurable quality modes and comprehensive benchmarking infrastructure, enabling reproducible evaluation and facilitating adoption for bandwidth-constrained communication scenarios. The complete source code and online speech reconstruction demo is publicly available at \url{https://github.com/dywsy21/STCTS}.
\end{itemize}

Beyond bitrate efficiency, our modular architecture offers several advantages: flexible deployment support (scalable from edge devices to accelerated servers), privacy-preserving encryption (independent encryption of text, prosody, and speaker data), model upgradeability (drop-in replacement of STT/TTS components and text/timbre compression techniques as technology advances), and interpretable transmission (human-readable text enables debugging, logging, and secondary applications). These properties position STCTS as a versatile framework for diverse deployment scenarios including maritime communication, satellite IoT networks, tactical military systems, and large-scale voice social platforms.

\section{Related Works}

Our approach integrates speech factorization, compression, and synthesis technologies to achieve ultra-low bitrate voice communication while preserving naturalness and speaker identity. This section surveys prior work across six key areas: (1) low-bitrate speech coding (covering traditional, neural, and semantic codecs), (2) STT-TTS architectures for bandwidth-constrained scenarios, (3) speech disentanglement methods that decompose speech into content, prosody, and timbre, (4) Speech-to-Text(STT) systems (particularly streaming and multilingual models), (5) text compression techniques (including neural language models), and (6) expressive text-to-speech(TTS) synthesis systems (focusing on voice cloning and prosody control). While individual components from these domains have been explored separately, our key contribution lies in their integration and optimization for natural, expressive communication at ultra-low bitrates—a scenario that demands different design choices than prior work in IoT communication (which sacrifices expressiveness) or speech disentanglement (which requires frame-level transmission).

\subsection{Low-Bitrate Speech Coding}

\textbf{Traditional speech codecs.} Traditional speech codecs (e.g., Opus, EVS) can operate down to a few kb/s. For example, Opus supports bitrates down to $\approx$6 kb/s (wideband speech) \cite{opus}, while specialized parametric codecs like Codec 2 handle ultra-low rates (0.7--3.2 kb/s) \cite{opus}. However, below $\sim$10 kb/s the quality of waveform coders degrades rapidly \cite{skoglund2020}. In response, neural and hybrid codecs have been developed to extend the range of intelligible speech at very low rates. Early work used linear-prediction plus RNN vocoders: LPCNet \cite{lpcnet} codes speech at 1.6 kb/s in real time, yielding much higher quality than classic MELP. More recently, LSPNet \cite{lspnet} extends LPCNet to about 1.2 kb/s by encoding line-spectral pairs and employing a joint time-frequency loss; it reports quality superior to both traditional codecs and prior neural codecs at that rate.

\textbf{End-to-End Neural Codecs.} Modern neural audio codecs use deep autoencoders or generative models. For instance, SoundStream \cite{soundstream} uses a convolutional encoder and residual vector quantizer to compress audio at 3--18 kb/s; at 3 kb/s it outperforms Opus at 12 kb/s in subjective tests. EnCodec \cite{encodec} uses a multi-band transformer VQ-VAE to compress high-fidelity (24 kHz) audio; by quantizing its latent space, it reduces bitrate by $\approx$40\% with little loss in quality. Google's Lyra V2 \cite{lyra} builds on SoundStream to deliver 3.2/6/9.2 kb/s modes for voice; at 6 kb/s it outperforms standard telco codecs (EVS/AMR-WB) and matches Opus quality while using only $\sim$50--60\% of the bandwidth. MLow \cite{mlow} is a low-complexity CELP-based codec optimized for 6 kb/s, reportedly doubling the perceptual quality of Opus at that rate (POLQA MOS $\approx$3.9 vs. 1.9) with $\sim$10\% lower compute.

\textbf{Hybrid and Other Neural Codecs.} Hybrid schemes combine parametric encoders with neural decoders. For example, Skoglund \& Valin \cite{skoglund2020} proposed decoding Opus parameters (6 kb/s) using neural vocoders: a listening test showed that synthesizing Opus 6 kb/s with LPCNet produced far better quality than Opus's standard decoder. Other work has explored GAN or diffusion-based codecs (e.g., AudioDec, DAC) but these require higher complexity. In the hybrid category, LPCNet and LSPNet (above) are prominent, as are deep neural networks for phase-aware speech enhancement \cite{hasannezhad2022} and RNNoise for noise suppression, and WaveRNN-based vocoders.

\textbf{Semantic/Generative Compression.} Recent research explores encoding high-level speech content rather than raw waveform. For instance, SemantiCodec \cite{semanticodec} uses a transformer-based semantic encoder (AudioMAE features) plus an acoustic residual, compressing diverse audio (speech, music, SFX) into $<$100 tokens/sec ($<$1 kb/s) while preserving quality. Similarly, Collette et al. \cite{collette2024} propose a semantic compression using generative voice models to factor speech into content and style; their method achieves perceptual quality beyond EnCodec at $\sim$650 bps. These approaches suggest future codecs may transmit text-like representations (or semantic features) and reconstruct speech with high fidelity.

\subsection{STT-TTS Architectures for Low-Bandwidth Communication}

The concept of using STT-TTS pipelines for bandwidth-constrained communication has been explored in IoT and satellite communication domains. Urazayev et al. \cite{urazayev2024voice} proposed a voice communication system for LoRaWAN networks that transmits text transcriptions over low-data-rate IoT channels (typically 0.3--5 kbps). Their approach achieves significant bandwidth reduction by converting speech to text at the sender, transmitting the compressed text, and synthesizing speech at the receiver using TTS. However, their work focuses primarily on IoT-specific challenges such as LoRaWAN protocol integration and network reliability, without explicit modeling of prosody or speaker identity. The synthesized speech uses generic voices without speaker adaptation, limiting its applicability to scenarios where speaker recognition and expressive communication are important.

Other work in satellite and tactical communication has explored similar STT-TTS architectures for bandwidth savings. These systems typically prioritize intelligibility and robustness over naturalness, often sacrificing prosodic expressiveness and speaker characteristics to minimize bitrate. While effective for mission-critical communications where only semantic content matters, such approaches are less suitable for general-purpose voice communication where users expect to recognize speakers and perceive emotional nuances.

Unlike prior STT-TTS systems that focus solely on semantic content transmission, our work explicitly models and transmits prosodic features and speaker embeddings alongside text. This enables us to preserve not just \textit{what} is said, but also \textit{how} it is said and \textit{who} is speaking. Furthermore, we introduce novel compression strategies tailored to each component: context-aware text compression, sparse prosody transmission with TTS interpolation, and amortized speaker embedding transmission. These innovations allow us to achieve ultra-low bitrates ($\sim$80 bps) while maintaining near-natural quality and speaker fidelity—a capability absent in prior STT-TTS systems designed for IoT or tactical scenarios.

\subsection{Speech Disentanglement}

Speech disentanglement aims to decompose speech signals into independent representations corresponding to different factors of variation. Recent work has explored separating speech into content (text), timbre (speaker identity), and prosody components.

\textbf{End-to-End Disentanglement Methods.} SpeechTripleNet \cite{speechtriplenet} and recent zero-shot approaches \cite{guo2024} propose an end-to-end framework for disentangling speech into three orthogonal representations: linguistic content, speaker timbre, and prosodic information. The model uses adversarial training with three discriminators to ensure that each representation captures only its intended factor while remaining invariant to others. SpeechSplit \cite{speechsplit} similarly decomposes speech into content, rhythm, pitch, and timbre using an autoencoder architecture with carefully designed information bottlenecks. These methods demonstrate that explicit factorization improves controllability in speech synthesis and voice conversion tasks.

\textbf{Self-Supervised Disentanglement.} More recent approaches leverage self-supervised learning to learn disentangled representations without explicit supervision. For instance, VQMIVC \cite{vqmivc} uses vector quantization to enforce discrete content representations while learning continuous speaker embeddings, achieving high-quality voice conversion. DiscreTalk \cite{discretalk} extends this by incorporating prosody modeling through separate prosody encoders, enabling fine-grained control over speaking style during synthesis.

\textbf{Comparison with Our Approach.} While speech disentanglement methods share the goal of factorizing speech into content, prosody, and timbre, they differ fundamentally from our work in objective and design. Disentanglement methods are primarily concerned with \textit{learning unsupervised representations} through adversarial training or self-supervised objectives, aiming to achieve clean separation of factors for downstream tasks like voice conversion, emotion transfer, or controllable synthesis. In contrast, our system operates in the \textit{component-wise compression and transmission} domain, where the goal is to minimize bitrate while preserving perceptual quality. We leverage \textit{off-the-shelf pre-trained models} (STT, prosody extractors, speaker embedding networks) rather than learning disentangled representations from scratch. This design choice offers several advantages: (1) modularity—components can be independently upgraded as better models emerge; (2) interpretability—transmitted text is human-readable, enabling logging and debugging; (3) efficiency—we exploit domain-specific compression strategies (e.g., Context-aware compression for text, sparse keyframe transmission for prosody) that would be difficult to integrate into end-to-end learned representations.

Furthermore, our prosody transmission strategy differs significantly from disentanglement methods. While prior work encodes prosody as continuous latent vectors or discrete tokens that must be transmitted at frame-level rates ($\sim$50--100 Hz), we exploit the \textit{interpolation capability of modern TTS models} to transmit prosody at extremely sparse rates (0.1--1 Hz), reducing prosody bitrate to $<$14 bps—orders of magnitude lower than what frame-level transmission would require. This sparse transmission strategy is enabled by our recognition that conversational prosody varies smoothly over multi-second spans, allowing TTS models to reconstruct natural prosody contours from sparse keyframes.

\subsection{Speech-To-Text (STT) Systems}

For speech-to-text in real time or low-resource settings, modern ASR models typically use powerful neural architectures, often with pre-training or streaming optimizations. OpenAI's Whisper \cite{whisper} is a large encoder--decoder Transformer trained on 680k hours of multilingual audio; it supports ASR, translation, language ID, etc. across $\sim$100 languages. Whisper is highly robust to noise and accents and often outperforms specialized models on many benchmarks. Conformer \cite{conformer} augments the Transformer with convolutional modules to capture local and global context; it achieved 2.1\%/4.3\% WER on LibriSpeech without an external language model, setting a new state of the art.

\textbf{Pretrained Self-Supervised Models.} Large SSL speech models (Wav2Vec 2.0, HuBERT, WavLM, etc.) are often fine-tuned for ASR. Delétang et al. note that models like Wav2Vec2, HuBERT, WavLM and Meta's MMS provide strong predictive performance but require task-specific fine-tuning \cite{deletang2024}. These models enable ASR in low-resource scenarios by transferring knowledge from large unlabeled datasets.

\textbf{Streaming/Realtime Architectures.} For on-device or low-latency ASR, streaming RNN-Transducer or Conformer-Transducer models are used (e.g., Emformer, optimized Conformer). These allow incremental inference with limited lookahead. In practice, hybrid approaches combine small neural LM or CTC models with streaming encoders.

\subsection{Existing Compression Techniques}

Existing compression techniques, especially text compression, typically uses entropy coding (Huffman, arithmetic or ANS) on top of a language model. Shannon's source-coding theorem implies an optimal code length of $-\log P(\text{token})$ bits; in practice one can feed LM-predicted probabilities into arithmetic coding for near-optimal compression \cite{deletang2024}. For example, Delétang et al. note that lossless compression with a probabilistic model can be achieved by Huffman, arithmetic or ANS coding. In short, we would tokenize (e.g., wordpieces) and then apply arithmetic coding driven by a language model trained on transcripts.

Recent works show that large neural LMs vastly outperform classic compressors. Language Modeling is Compression \cite{deletang2024} demonstrated that a 70B-parameter Transformer (Chinchilla) can compress LibriSpeech to $\sim$16.4\% of raw size -- dramatically better than FLAC (30.3\%). LMCompress \cite{lmcompress} uses similar ideas across data types: it ``shatters all previous compression records,'' achieving text compression at roughly one-third the size of the prior best text compressor (zpaq). These results imply that an off-the-shelf ASR model (if treated as a language model) could serve as a near-optimal text compressor. In practice, we would likely use a smaller LM or on-device model to balance speed and size. In summary, entropy coding guided by LMs yields the state of the art: Huffman/arithmetic coding on tokens using an ASR or NLP model would minimize the bits needed for transcripts. However, since its actual complete implementation is not publicly released yet and taking into account the potential limitations of computational resources (a tradeoff between Real Time Factor / RTF and Bitrate), as of now we do not employ LM-based text compression techniques.

\subsection{Expressive Text-to-Speech (TTS) Systems}

A wide range of modern TTS models support expressive, speaker-specific synthesis from text plus prosody or embedding inputs. Notable examples include:

\textbf{Tacotron 2.} \cite{tacotron2} An attention-based seq2seq model that predicts mel-spectrograms from text, followed by a neural vocoder. It ``synthesizes speech with Tacotron-level prosody and WaveNet-level audio quality,'' achieving near-human sound quality. Tacotron2 requires a trained vocoder (e.g., WaveNet or HiFi-GAN) but produces very natural prosody. Furthermore, adaptation techniques \cite{bollepalli2019} have demonstrated the importance of modeling prosodic features like duration and energy for expressive synthesis.

\textbf{FastSpeech 2.} \cite{fastspeech2} A non-autoregressive Transformer-based TTS that conditions directly on duration, pitch, and energy extracted from speech. By training with ground-truth durations/intonation, FastSpeech2 avoids alignment issues and speeds up synthesis. It achieves 3$\times$ faster training and higher quality than original FastSpeech, even surpassing many autoregressive models. This makes it well-suited for real-time use.

\textbf{XTTS.} \cite{xtts} A recent zero-shot multi-speaker TTS. Building on the Tortoise architecture, it is trained on 16 languages and can synthesize new voices and languages without fine-tuning. XTTS achieves state-of-the-art cross-lingual voice cloning performance in most of those languages. It demonstrates that massively multilingual, zero-shot cloning is feasible with large models.

\textbf{Zonos.} \cite{zonos} An open-weight 1.6B model suite (Transformer and SSM hybrid) trained on $\sim$200k h of speech (English plus Chinese, Japanese, etc.). Zonos produces highly expressive, natural speech from text given a speaker embedding or audio example. It enables high-fidelity voice cloning from just 5--30 s of reference audio, and even allows control over speaking rate, pitch, and emotions (sadness, anger, etc.). The creators report that Zonos' quality matches/exceeds top proprietary TTS, and it outputs 44 kHz speech.

In all these TTS systems, one can provide prosody control signals (pitch/energy), and/or a learned speaker embedding (or reference audio) to capture the desired voice and style. Together with a high-quality vocoder (e.g., HiFi-GAN), these models can reconstruct speech that preserves the speaker's identity and expressiveness. This feature will prove crucial in our work.

\section{Methodology}

We propose an end-to-end voice communication system that achieves ultra-low bitrate transmission ($\sim$80 bps) by leveraging semantic compression through the STCTS pipeline. Unlike traditional waveform or neural audio codecs that operate in the acoustic domain, our approach transforms speech into a compressed semantic representation (text) with auxiliary prosodic and speaker information, then reconstructs natural-sounding speech at the receiver. This represents a $75\times$ reduction compared to standard Opus codec ($\sim$6 kbps) and $12\times$ reduction compared to state-of-the-art neural codecs like EnCodec ($\sim$1 kbps) while preserving audio quality and speaker characteristics. This section details the system architecture, individual components, and implementation choices.\footnote{Parameters marked with * are configurable via our custom defined YAML configuration files. Values shown correspond to the balanced mode unless otherwise specified.}

\subsection{System Overview}

Our system consists of three main stages operating in a duplex communication channel:

\textbf{Stage 1: Speech Analysis and Encoding.} In this stage, the text, prosody and timbre information are extracted from the sender's audio. The sender's audio is processed through Voice Activity Detection (VAD) to filter out silence periods, followed by Speech-to-Text (STT) conversion to extract linguistic content. Simultaneously, prosody extraction captures intonation, rhythm, and speaking style, while speaker embedding extraction encodes voice timbre characteristics.

\textbf{Stage 2: Compression and Transmission.} The extracted features are compressed using appropriate algorithms: text compression via Brotli with optional preprocessing and context-aware compression, prosody quantization with delta encoding, and speaker embedding transmission with cache and change detection. These compressed packets are transmitted through a WebRTC data channel with priority-based queuing.

\textbf{Stage 3: Decompression and Speech Reconstruction.} The receiver decompresses the data stream and reconstructs speech through text decompression, prosody reconstruction with interpolation for missing frames, speaker-conditioned TTS synthesis, and audio playback with quality enhancement.

\subsection{Speech-to-Text Module}

Our STT pipeline is designed to achieve real-time transcription with minimal latency while maintaining robustness across diverse acoustic environments. The system combines a state-of-the-art multilingual model with voice activity detection and streaming processing architecture to enable responsive speech recognition suitable for interactive communication.

\subsubsection{Model Selection}

Our modular architecture supports drop-in replacement of STT models, enabling users to select engines optimized for their specific requirements (latency, accuracy, language coverage, or computational constraints). For the baseline implementation, we select FasterWhisper (small model*) based on its balanced performance across multiple criteria critical for real-time communication: (1) \textbf{computational efficiency}—achieving low Real-Time Factor (RTF) on modern hardware, allowing us to trade computational power for bandwidth savings; (2) \textbf{robustness}—maintaining WER $<$10\% on LibriSpeech under realistic noise conditions (SNR $>$10 dB) due to training on 680k hours of diverse audio; (3) \textbf{multilingual support}—covering 100+ languages without language-specific models, essential for global deployment; and (4) \textbf{streaming compatibility}—supporting chunked inference with minimal lookahead ($<$500ms), critical for interactive latency requirements. The CTranslate2-optimized implementation provides 3--4$\times$ speedup over vanilla Whisper while preserving accuracy within 0.5\% WER \cite{whisper}. Alternative models (e.g., Wav2Vec2 for low-resource languages, or Conformer-Transducer for minimal latency) can be substituted without modifying downstream components, as the system interfaces solely through transcribed text output.

\subsubsection{Voice Activity Detection (VAD)}

To minimize bandwidth consumption and computational overhead, we implement Silero VAD \cite{silerovad} to detect speech segments. The VAD operates on 30ms audio frames with a speech probability threshold of 0.5*, minimum speech duration of 250ms to filter out false positives, and minimum silence duration of 500ms before considering a speech segment complete. This configuration reduces unnecessary processing of silence periods while maintaining responsiveness to natural speech patterns.

\subsubsection{Streaming Architecture}

The STT module processes audio in overlapping windows to enable low-latency transcription. We buffer audio chunks of 400ms* with 50ms overlap between consecutive windows. This design allows the system to begin transcription before a complete utterance finishes, maintain context across chunk boundaries through overlap, and achieve end-to-end latency under 500ms* from speech to text output.

To improve transcription quality in the streaming context, we implement a minimum buffer threshold of 25 audio chunks ($\sim$250ms) before initiating STT processing. Additionally, we enforce a minimum transcription length of 3 characters to filter out spurious detections from background noise.

Alternatively, the system supports a push-to-talk mode similar to walkie-talkie operation. In this mode, users activate a button to begin speaking, and audio is buffered locally until the button is released. The complete utterance is then transcribed and transmitted as a single packet, reducing network overhead and enabling more aggressive compression through larger context windows. This mode is particularly suitable for half-duplex communication scenarios or when network conditions require minimizing packet fragmentation.

\subsection{Prosody and Timbre Extraction}
\label{sec:prosody_timbre}

Beyond linguistic content, natural speech communication depends critically on prosodic cues (intonation, rhythm, emphasis) and speaker identity. Our system extracts compact representations of these paralinguistic features to enable expressive and personalized voice reconstruction at bitrates far lower than acoustic encoding would require.

\subsubsection{Prosody Feature Extraction and Encoding}

Prosodic information enables expressive communication while maintaining ultra-low bitrate through sparse transmission. We formalize the complete prosody encoding pipeline from raw audio to compressed bitstream.

\textbf{Feature Extraction.} Given input audio $\mathbf{x}[n]$ sampled at 16 kHz, we extract three prosody features at frame rate $f_p = 100$ Hz (10ms frame shift):

\textit{Pitch Contour:} Fundamental frequency $F_0[t]$ is extracted via YIN \cite{yin} ($F_0=0$ for unvoiced). We log-scale and normalize $F_0$ using speaker statistics $\mu_{F_0}, \sigma_{F_0}$ from the first 3 seconds:
\begin{equation}
\hat{F}_0[t] = \begin{cases}
\frac{\log(F_0[t]) - \mu_{F_0}}{\sigma_{F_0}} & \text{if } F_0[t] > 0 \\
0 & \text{otherwise}
\end{cases},
\end{equation}
where $\hat{F}_0[t]$ is the normalized log-pitch at frame $t$, $\mu_{F_0}$ is the mean log-pitch, and $\sigma_{F_0}$ is the standard deviation of the log-pitch. This normalization ensures speaker-independent representation and concentrates values around zero for efficient quantization.

\textit{Energy Envelope:} RMS energy $E[t]$ is computed over 40ms windows ($N_w = 640$ samples):
\begin{equation}
E[t] = \sqrt{\frac{1}{N_w} \sum_{i=0}^{N_w-1} x[tN_s + i]^2},
\end{equation}
where $x[n]$ is the input audio signal, $N_w$ is the window size, and $N_s = 160$ is the hop size. Energy is normalized to the speaker's log-energy dynamic range:
\begin{equation}
\hat{E}[t] = \frac{\log(E[t] + \epsilon) - \log(E_{\text{min}})}{\log(E_{\text{max}}) - \log(E_{\text{min}})},
\end{equation}
where $\hat{E}[t]$ is the normalized energy, $E_{\text{min}}$ and $E_{\text{max}}$ are the 5th and 95th percentiles of the speaker's energy distribution (computed over a 10-second sliding window), and $\epsilon = 10^{-6}$ prevents log-domain singularities. These acoustic features have been shown to effectively capture emotional content across languages \cite{li2019}.

\textit{Speaking Rate:} We estimate instantaneous speaking rate $R[t]$ (syllables/second) through syllable nucleus detection. We apply a bandpass filter (300--3000 Hz) to extract the speech envelope, detect local maxima exceeding an adaptive threshold, and count nuclei within a 1-second centered window:
\begin{equation}
R[t] = \frac{1}{T_w} \sum_{\tau=t-T_w/2}^{t+T_w/2} \mathbbm{1}[\text{nucleus detected at } \tau],
\end{equation}
where $T_w = 1$ second is the window duration, and $\mathbbm{1}[\cdot]$ is the indicator function which is 1 if a nucleus is detected at time $\tau$ and 0 otherwise. Speaking rate is then normalized relative to the speaker's baseline rate $\mu_R$ (typically 3--5 syllables/sec for conversational speech):
\begin{equation}
\hat{R}[t] = \frac{R[t] - \mu_R}{\sigma_R},
\end{equation}
where $\hat{R}[t]$ is the normalized speaking rate, $\mu_R$ is the mean speaking rate, and $\sigma_R$ is the standard deviation of the speaking rate.

\subsubsection{Timbre as Speaker Embedding}

To preserve speaker identity, we extract a 192-dimensional* speaker embedding using the ECAPA-TDNN model* \cite{ecapatdnn} from SpeechBrain \cite{speechbrain}. This embedding is computed once at call initialization and updated only when speaker change is detected* with a cosine similarity threshold $<$ 0.7* from the previous embedding. The embedding is quantized to float16 precision*, resulting in a 384-byte payload per transmission.

\subsubsection{Timbre Transmission Strategy}

We employ an amortized transmission strategy for the 384-byte speaker embeddings. The full embedding is sent only at call start or upon speaker change. For a 45-second utterance, this amortizes to $\sim$68 bps, dropping to 5--20 bps in longer calls.

To further optimize, the receiver caches timbre profiles locally. The sender transmits a lightweight TIMBRE\_PROFILE packet (4--8 bytes) if the speaker is already cached, sending the full embedding only for new or significantly changed voices. This mechanism effectively reduces timbre overhead to near-zero for recurring speakers.

\subsection{Compression Pipeline}

The compression stage transforms extracted features into a minimal bitstream suitable for bandwidth-constrained transmission. We employ component-specific strategies tailored to each data type: semantic-aware compression for text, temporal delta coding for prosody, and amortized transmission, caching and universal compression techniques for speaker characteristics.

\subsubsection{Text Compression}

Transcripts are compressed using Brotli* (level 5*), yielding $\sim$70 bps. We enhance this with context-aware optimization, maintaining a sliding window to build a dynamic dictionary for conversation-specific terms. This adaptive approach improves encoding efficiency for recurring jargon or names. Optional preprocessing (filler removal, abbreviations, punctuation minimization) further reduces text volume by 5--10\% with minimal naturalness loss.

\subsubsection{Prosody Compression}

The prosody stream is transmitted sparsely to minimize bandwidth consumption. Rather than sending continuous prosody features, we transmit prosody updates only when significant changes occur or at keyframe intervals (0.5 Hz*, every 2 seconds). In our minimal mode, prosody updates are sent as infrequently as 0.1 Hz, resulting in only 2--4 bytes of prosody data over a 45-second utterance. The receiver interpolates prosody between sparse updates*, maintaining naturalness while achieving almost negligible prosody bitrate (less than $\sim$14 bps).

When prosody is transmitted, it undergoes three-stage compression inspired by traditional parametric coding approaches \cite{lpcnet}:

\begin{figure*}[t]
\centering
\includegraphics[width=\textwidth]{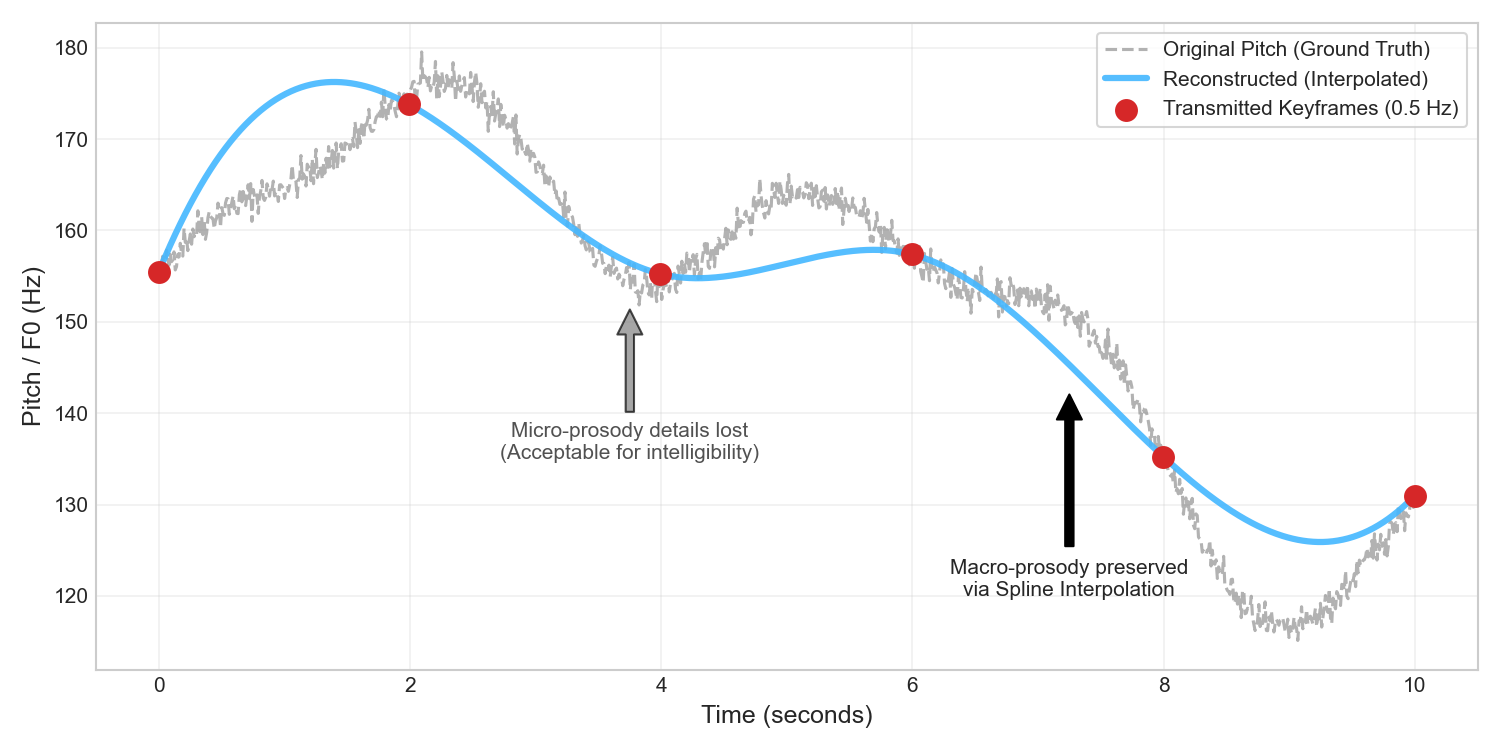}
\caption{Sparse prosody interpolation principle. The system transmits prosody keyframes at a very low rate (e.g., 0.5 Hz, red dots). The receiver reconstructs the continuous pitch contour (blue line) via cubic spline interpolation, which closely approximates the original macro-prosody (gray dashed line) while discarding micro-jitter, achieving ultra-low prosody bitrate ($<$14 bps).}
\label{fig:prosody_interpolation}
\end{figure*}

\textbf{Sparse Sampling and Delta Encoding.} Rather than transmitting prosody at the native 100 Hz extraction rate, we employ sparse keyframe sampling at rate $f_k$ (configurable: 0.1--1 Hz). Let $\mathbf{p}[t] = [\hat{F}_0[t], \hat{E}[t], \hat{R}[t]]^\top$ denote the normalized prosody vector. We select keyframes at indices $\mathcal{T}_k = \{t_0, t_0 + \Delta t, t_0 + 2\Delta t, \ldots\}$ where $\Delta t = \lfloor f_p / f_k \rfloor$. For keyframes $t \in \mathcal{T}_k$, we compute temporal deltas:
\begin{equation}
\Delta \mathbf{p}[t] = \mathbf{p}[t] - \mathbf{p}[t - \Delta t].
\end{equation}
The first keyframe transmits absolute values: $\Delta \mathbf{p}[t_0] = \mathbf{p}[t_0]$.

\textbf{Non-Uniform Quantization.} Delta values $\Delta \mathbf{p}[t]$ are quantized using a dead-zone uniform quantizer tailored to the sparse nature of the signal. For pitch deltas $\Delta \hat{F}_0[t]$ (quantized to $b_F$ bits*, e.g., 6 bits):
\begin{equation}
\Delta \hat{F}_0^{(q)}[t] = \begin{cases}
0 & \text{if } |\Delta \hat{F}_0[t]| < \tau_F \\
\text{sign}(\Delta \hat{F}_0[t]) \cdot \lceil |\Delta \hat{F}_0[t]| / \alpha_F \rceil & \text{otherwise}
\end{cases},
\end{equation}
where $\tau_F = 0.05$ is a dead-zone threshold (suppressing imperceptible changes), and $\alpha_F$ is the quantization step size determined by the bit budget. This scheme efficiently captures significant prosodic shifts while ignoring minor fluctuations. Energy and speaking rate follow analogous quantization with $b_E = 5$ bits* and $b_R = 5$ bits*, respectively.

\textbf{Entropy Coding and Packetization.} Quantized delta vectors are entropy-coded using Huffman coding, which exploits the non-uniform distribution of prosody deltas (with strong concentration near zero). Each keyframe packet contains:
\begin{multline}
\text{Packet}_{\text{prosody}}(t) = [\text{timestamp}(t), \\
\text{Huffman}(\Delta \hat{F}_0^{(q)}[t], \Delta \hat{E}^{(q)}[t], \Delta \hat{R}^{(q)}[t])].
\end{multline}
For $f_k = 0.5$ Hz (balanced mode), this yields $\sim$16--20 bits per keyframe, resulting in 8--10 bps prosody bitrate. The receiver reconstructs continuous prosody at 100 Hz through cubic spline interpolation between received keyframes.

\subsubsection{Timbre Compression}

Speaker embeddings (192-dimensional vectors) are first quantized to lower precision (float16 or float32) to reduce the baseline payload size. To further minimize bandwidth, we apply universal lossless compression algorithms (zlib or Brotli) to the quantized byte stream. Since speaker embeddings often exhibit statistical redundancies, this additional compression step typically yields a 10--20\% reduction in payload size without any loss of information beyond the initial quantization. This ensures that the critical speaker identity information is transmitted as efficiently as possible.

\subsection{Network Transport and Reliability}
\label{sec:transport}

At ultra-low bitrates ($\sim$80 bps), we employ a priority-based packet transmission strategy where each data type (TEXT, PROSODY, TIMBRE) is assigned a priority level reflecting its perceptual importance. The transport layer employs minimal packet headers (4--8 bytes) to reduce protocol overhead to below 10\% of total bandwidth for typical packet sizes.

\textbf{Differentiated Reliability and Graceful Degradation.} We apply different reliability guarantees tailored to each stream's tolerance for loss:

\begin{figure*}[t]
\centering
\includegraphics[width=\textwidth]{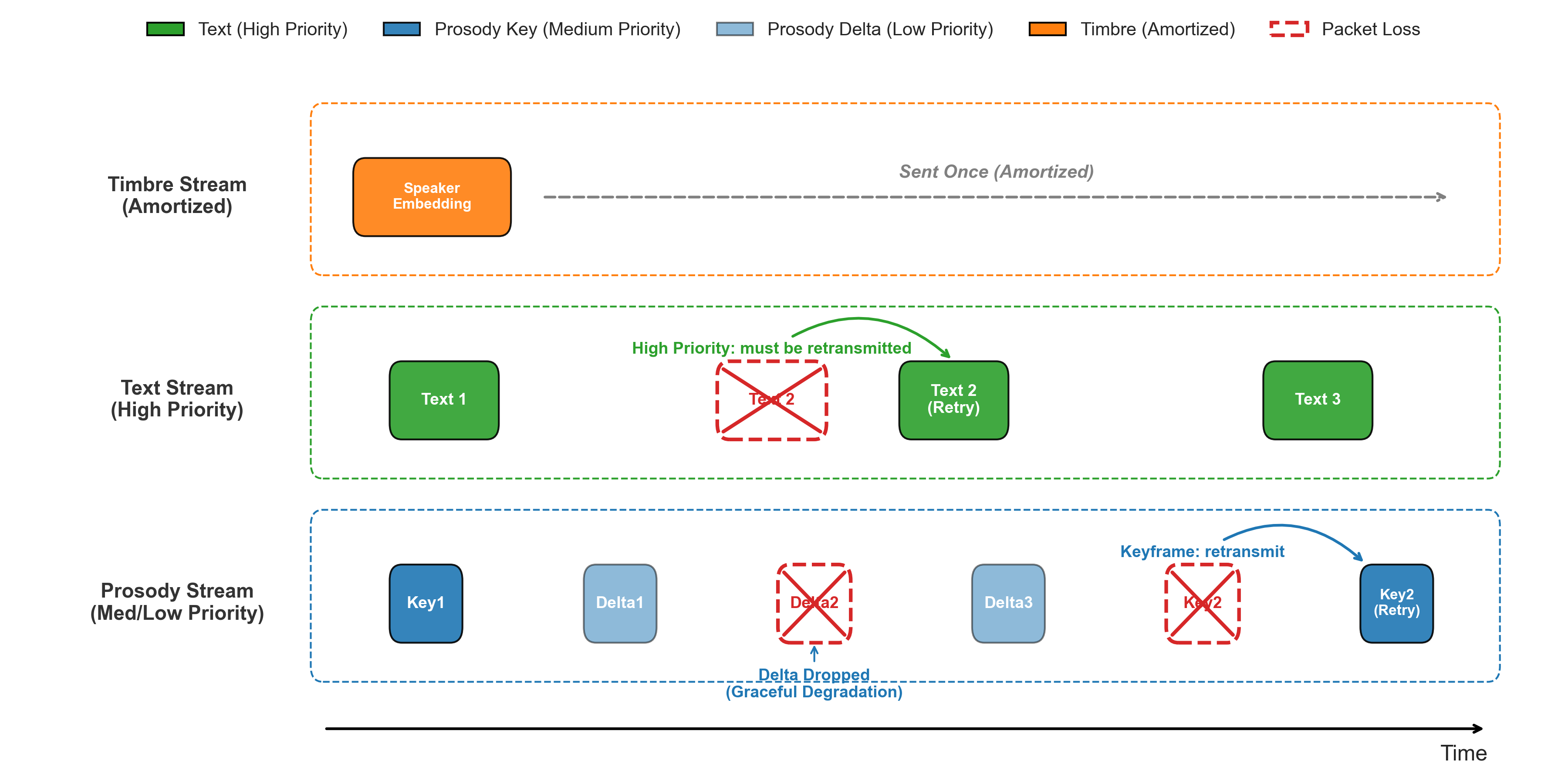}
\caption{Prioritized transmission and error handling mechanism. Text packets (High Priority) are retransmitted upon loss to ensure semantic integrity. Prosody Keyframes (Medium Priority) are retransmitted to maintain interpolation anchors, while Prosody Deltas (Low Priority) are discarded if lost, allowing graceful degradation. Timbre packets are sent once per speaker and cached, amortizing bandwidth cost.}
\label{fig:transmission_mechanism}
\end{figure*}

\begin{itemize}
\item \textit{TEXT} (HIGH priority): Requires absolute integrity due to the cascading failure mode of entropy coding—a single corrupted byte renders the entire compressed block undecompressable. Failed TEXT packets trigger immediate retransmission.
\item \textit{TIMBRE} (HIGH priority): Speaker embeddings are critical for identity preservation but transmitted infrequently (once per speaker, cached at the receiver once received). Loss and cache miss simultaneously occurring triggers retransmission.
\item \textit{PROSODY keyframes} (MEDIUM priority): Sparse prosody updates are retransmitted once if lost, as interpolation quality degrades significantly with missing keyframes.
\item \textit{PROSODY deltas} (LOW priority): Best-effort delivery without retransmission. The receiver interpolates through missing deltas with graceful degradation.
\end{itemize}

This tiered approach ensures that critical semantic information (text and speaker identity) maintains high reliability while tolerating graceful degradation in prosodic expressiveness under severe packet loss, consistent with the perceptual robustness to be demonstrated in our evaluation.

\subsection{Text-to-Speech Synthesis}

The receiver reconstructs natural-sounding speech by conditioning a neural TTS model on the transmitted text, prosody, and speaker features. Our synthesis pipeline emphasizes voice fidelity and expressive control while maintaining real-time performance.

\subsubsection{Model Selection}

We select Coqui XTTS-v2 \cite{xtts} for its balance of capabilities:

\textbf{(1) Zero-Shot Cloning:} High speaker similarity ($>0.85$) from just 3 seconds of audio.

\textbf{(2) Explicit Prosody Conditioning:} Direct pitch/energy control via cross-attention, essential for our sparse prosody stream.

\textbf{(3) Multilingual Support:} Covers 16 languages, matching our STT frontend.

\textbf{(4) Streaming Synthesis:} Enables real-time performance (RTF $\sim$0.4) with HiFi-GAN vocoding.

This combination supports our low-bandwidth, high-expressiveness goals. Future work may explore newer models like StyleTTS-2.

\subsubsection{Prosody Conditioning}

Reconstructed prosody features are injected into the TTS model at multiple stages, following expressive TTS paradigms \cite{tacotron2, fastspeech2}. Pitch contours modulate the fundamental frequency of the generated mel-spectrogram, energy envelopes control the loudness of each frame, and speaking rate modulates the pace of token generation. This explicit conditioning ensures that the synthesized speech reflects the original speaker's expressive patterns.

\subsubsection{Speaker Conditioning}

The received speaker embedding serves as the reference for voice cloning. XTTS-v2 conditions its generation on this embedding through cross-attention mechanisms in the decoder. When the embedding is updated (speaker change or periodic refresh), the synthesizer adapts to the new voice characteristics within 1--2 seconds.

\subsection{Quality Modes}

To accommodate varying network conditions, we define three operational modes (Minimal, Balanced, and High Quality) with measured bitrates. Each mode is specified via a YAML configuration file that controls all system parameters, enabling flexible customization beyond the predefined profiles. In particular, the prosody sampling rate is a critical parameter that largely determines the trade-off between bitrate and quality; we determine its optimal values for these modes based on the comprehensive analysis presented in Section~\ref{sec:prosody_analysis}. Detailed specifications for each mode are provided in Appendix~\ref{app:quality_modes}.

\subsection{Implementation Details}

The system is implemented in Python 3.11 using FasterWhisper for STT, Silero VAD for voice activity detection, SpeechBrain for speaker embeddings, XTTS for synthesis, aiortc for WebRTC communication, and Brotli for text compression.

The sender and receiver run as asynchronous processes using Python's asyncio framework. Audio capture uses PyAudio with a 16 kHz sampling rate*, 16-bit depth, and 20ms frame size. The audio buffer chunk size* (1024 samples by default) and channel count* (mono by default) can be adjusted for different hardware configurations. The STT module processes audio in a dedicated thread pool to avoid blocking the main event loop. Similarly, TTS synthesis runs in a separate process to maintain real-time responsiveness.

A signaling server facilitates peer discovery and WebRTC connection establishment. The server is implemented using WebSockets and handles peer registration, session negotiation, and ICE candidate exchange. Once the WebRTC connection is established, all voice data bypasses the signaling server and flows directly peer-to-peer.

\section{Experiments}

We conduct comprehensive experiments to evaluate our STCTS system across multiple dimensions: bitrate efficiency, transcription accuracy, voice identity preservation, perceptual speech quality, noise resilience and computational efficiency. We first conduct \textit{prosody sampling rate analysis} to determine the optimal prosody sampling rate for our three quality modes, and then our evaluation compares three quality modes (minimal, balanced, and high-quality) against baseline codecs (Opus and EnCodec) and the Vevo framework \cite{collette2024} using standardized metrics and a large-scale speech corpus.

\subsection{Experimental Setup}

\subsubsection{Dataset}

We evaluate our system on the LibriSpeech corpus \cite{librispeech}, a widely-used benchmark for speech recognition research. LibriSpeech contains approximately 1,000 hours of read English speech derived from audiobooks in the LibriVox project, carefully segmented and aligned at 16 kHz sampling rate. The corpus is partitioned into multiple subsets based on acoustic conditions and speaker characteristics.

For our experiments, we evaluate our setup and the baseline setups on the full \texttt{test-clean} subset, which contains high-quality recordings with minimal background noise and clear articulation. This subset provides a controlled environment for measuring system performance under ideal acoustic conditions. Each sample contains complete sentences or utterances ranging from 5 to 30 seconds in duration, spoken by diverse speakers (both male and female) with various accents and speaking styles. We report mean values across all samples along with standard deviations in order to ensure statistical reliability.

It is important to clarify that while STCTS is designed for conversational voice communication, our evaluation primarily assesses the \textit{reconstruction quality} of the compression pipeline (STT $\rightarrow$ Compression $\rightarrow$ TTS) rather than conversational dynamics (e.g., turn-taking latency, overlapping speech). Since the core challenge lies in reconstructing intelligible and expressive speech from ultra-low bitrate semantic representations, single-channel read speech from LibriSpeech provides a rigorous and standardized benchmark for this purpose. The conversational aspects are handled by the networking layer (WebRTC) and do not fundamentally alter the factorization and compression algorithm's performance characteristics. Therefore, evaluating on a high-quality read speech corpus allows us to isolate and precisely measure the fidelity of our semantic reconstruction approach.

The choice of \texttt{test-clean} allows us to isolate the compression artifacts and reconstruction quality from environmental noise factors. We separately evaluate noise resilience in Section~\ref{sec:benchmark} using augmented test sets with additive noise at various signal-to-noise ratios.

\subsubsection{Baseline Systems}

We compare our approach against three representative baseline systems:

\textbf{Opus (6 kbps).} A widely-deployed traditional codec operating at its lowest recommended bitrate for wideband speech. Opus uses SILK for speech and CELT for music, with hybrid packet loss concealment. This represents the state-of-the-art in traditional parametric coding.

\textbf{EnCodec (1 kbps).} A recent neural audio codec using multi-scale VQ-VAE architecture. EnCodec represents the current frontier in neural waveform compression, achieving significantly lower bitrates than traditional codecs while maintaining reasonable quality.

\textbf{Vevo Framework ($\sim$650 bps).} A semantic compression system proposed by Collette et al. \cite{collette2024} that similarly uses generative voice models to decompose speech into content and style components. Vevo achieves ultra-low bitrates ($\sim$650 bps) comparable to our approach. However, since Vevo was evaluated on a different test set than ours, we report their published metrics as reference values only. These values provide context for our results but cannot be directly compared due to dataset differences. We mark Vevo results with an asterisk (*) in all tables to indicate this distinction.

\subsubsection{Evaluation Metrics}

We employ a comprehensive suite of metrics covering bitrate, intelligibility, speaker fidelity, and perceptual quality:

\textbf{Bitrate Metrics.} We measure the total transmitted bitrate in bits per second (bps), decomposed into three components:

\begin{itemize}
\item \textbf{Text, Prosody, Timbre Bitrates:} Individual component bitrates to analyze bandwidth allocation.
\item \textbf{Total Bitrate w/o Timbre:} Sustained bitrate excluding the one-time speaker embedding transmission, representing the long-term bandwidth consumption.
\end{itemize}

Bitrates are measured over the complete utterance duration including silence periods detected by VAD. For timbre, we amortize the one-time 384-byte embedding transmission over the utterance duration to compute an equivalent bitrate.

\textbf{STT Accuracy: Word Error Rate (WER).} We measure transcription accuracy using Word Error Rate, defined as:
\[
\text{WER} = \frac{S + D + I}{N},
\]
where $S$, $D$, $I$ are the number of substitutions, deletions, and insertions required to transform the recognized text into the reference transcription, and $N$ is the total number of words in the reference. Lower WER indicates better intelligibility. We compute WER by transcribing both the original and reconstructed audio using the same FasterWhisper model, then comparing the two transcriptions to isolate compression-induced errors.

\textbf{Speaker Similarity (SpkrSim).} We evaluate voice identity preservation by computing the cosine similarity between ECAPA-TDNN speaker embeddings \cite{ecapatdnn} extracted from the original and reconstructed audio:
\[
\text{SpkrSim} = \frac{\mathbf{e}_{\text{orig}} \cdot \mathbf{e}_{\text{recon}}}{\|\mathbf{e}_{\text{orig}}\| \|\mathbf{e}_{\text{recon}}\|},
\]
where $\mathbf{e}_{\text{orig}}$ and $\mathbf{e}_{\text{recon}}$ are the 192-dimensional speaker embeddings. Values above 0.85 typically indicate the same speaker according to standard equal error rate (EER) thresholds in speaker verification systems. This metric assesses how well the TTS synthesis preserves the original speaker's timbre and voice characteristics.

\textbf{Perceptual Evaluation of Speech Quality (PESQ).} PESQ \cite{pesq} is an ITU-T standard (P.862) for objective speech quality assessment in telecommunications. It compares the original and degraded signals in a perceptually-weighted frequency domain, producing scores from $-0.5$ to $4.5$ (higher is better). PESQ correlates well with subjective Mean Opinion Score (MOS) ratings, with scores above 2.5 considered acceptable telephony quality and above 3.5 considered excellent. We use the wideband (16 kHz) mode for all evaluations.

\textbf{Short-Time Objective Intelligibility (STOI).} STOI \cite{stoi} measures speech intelligibility by computing the normalized correlation between short-time segments of the original and processed signals in a temporal-frequency domain. It produces scores from 0 to 1, where higher values indicate better intelligibility. STOI has been validated to correlate strongly with human speech reception thresholds and is particularly effective for assessing intelligibility in noisy or distorted conditions.

\textbf{Non-Intrusive Speech Quality Assessment (NISQA).} Unlike PESQ and STOI which require reference audio (intrusive metrics), recent work has explored output-based assessment using autoencoders \cite{wang2019}, and NISQA \cite{nisqa} is a deep learning-based non-intrusive metric that predicts speech quality from the degraded signal alone. It produces a Mean Opinion Score (MOS) ranging from 1 to 5, where 5 represents excellent quality. NISQA additionally provides four quality dimensions: noisiness, coloration (spectral distortion), discontinuity (temporal artifacts), and loudness appropriateness. We report the overall MOS score as the primary quality indicator. NISQA's non-intrusive nature makes it particularly valuable for evaluating synthesis artifacts that may not be captured by intrusive metrics.

\subsubsection{Experiment Implementation Details}

All experiments are conducted on a system equipped with a single NVIDIA RTX 4080 GPU and an Intel Core i9-13900KF CPU.
For baseline comparisons, Opus encoding uses the \texttt{libopus} library with VBR mode disabled and bitrate constrained to 6 kbps. EnCodec uses the official implementation with the 24 kHz model quantized to 1 kbps (single codebook). Both baselines process the same test audio samples for fair comparison.

\subsection{Prosody Sampling Rate Analysis}
\label{sec:prosody_analysis}

We systematically investigate the relationship between prosody sampling rate and system performance to lay out the theoretical foundation for our three operational modes (minimal, balanced, and high-quality). Prosody sampling rate fundamentally determines the temporal granularity at which expressive features are transmitted, directly impacting both bandwidth consumption and reconstruction quality. Understanding this tradeoff is essential for informed configuration design.

We conduct a parameter sweep experiment on the LibriSpeech test-clean dataset, varying prosody sampling rate from 0.05 Hz (one update every 20 seconds) to 20 Hz (20 updates per second) while keeping all other parameters constant (small STT model, Brotli level 5, 192-dim float16 speaker embedding). For each sampling rate, we measure: (1) total bitrate (including timbre before amortization), (2) prosody bitrate contribution, (3) transcription accuracy (WER), (4) speaker similarity, (5) perceptual quality metrics (PESQ, STOI, NISQA), and (6) NISQA dimensional breakdown (noisiness, coloration, discontinuity, loudness).

\begin{figure*}[t]
\centering
\begin{subfigure}[b]{0.48\textwidth}
\centering
\includegraphics[width=\textwidth]{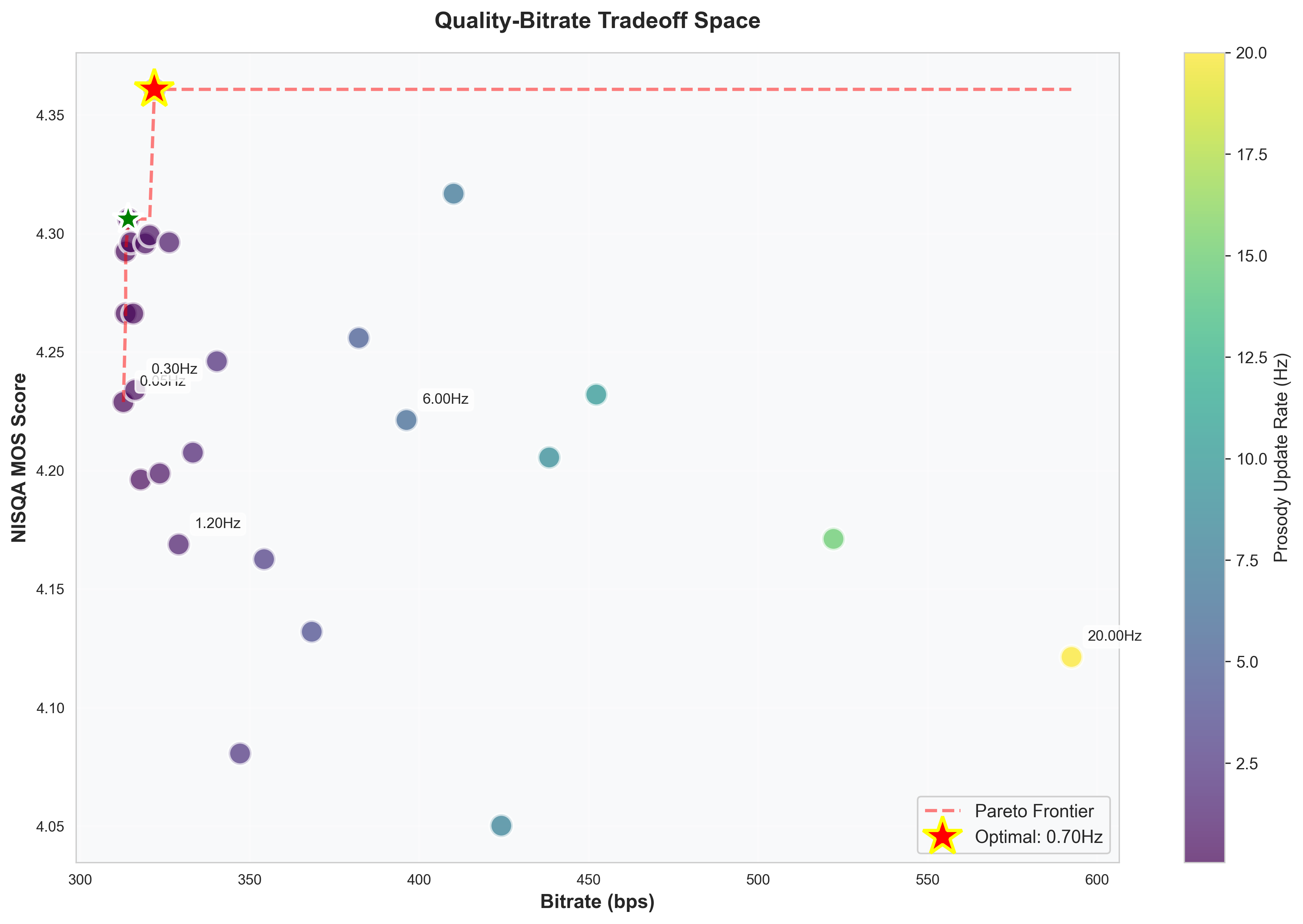}
\caption{Bitrate vs. quality tradeoff showing optimal operating regions.}
\label{fig:prosody_a}
\end{subfigure}
\hfill
\begin{subfigure}[b]{0.48\textwidth}
\centering
\includegraphics[width=\textwidth]{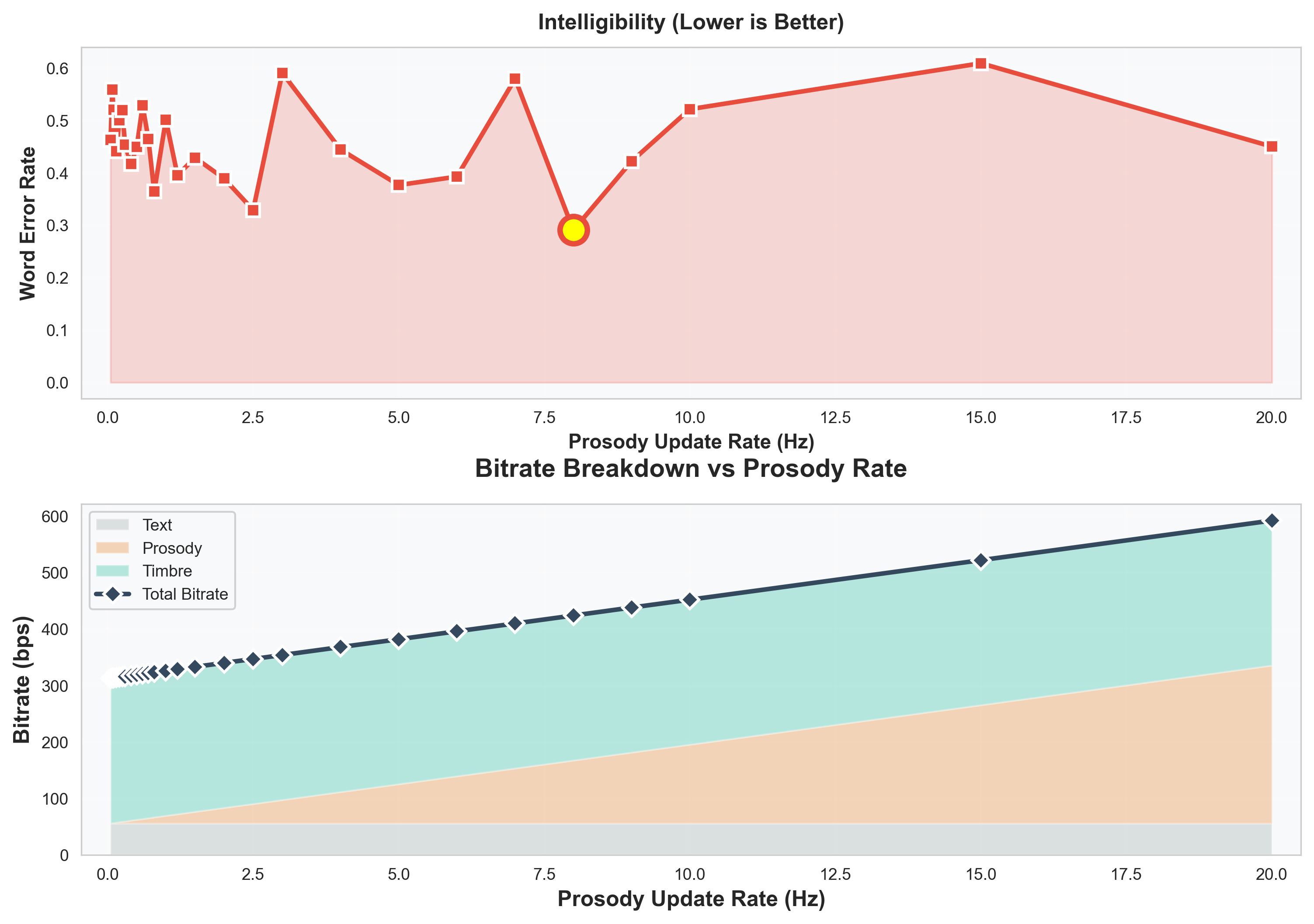}
\caption{Intelligibility metric (WER) across sampling rates and bitrate breakdown.}
\label{fig:prosody_b}
\end{subfigure}

\vspace{0.5em}

\begin{subfigure}[b]{0.48\textwidth}
\centering
\includegraphics[width=\textwidth]{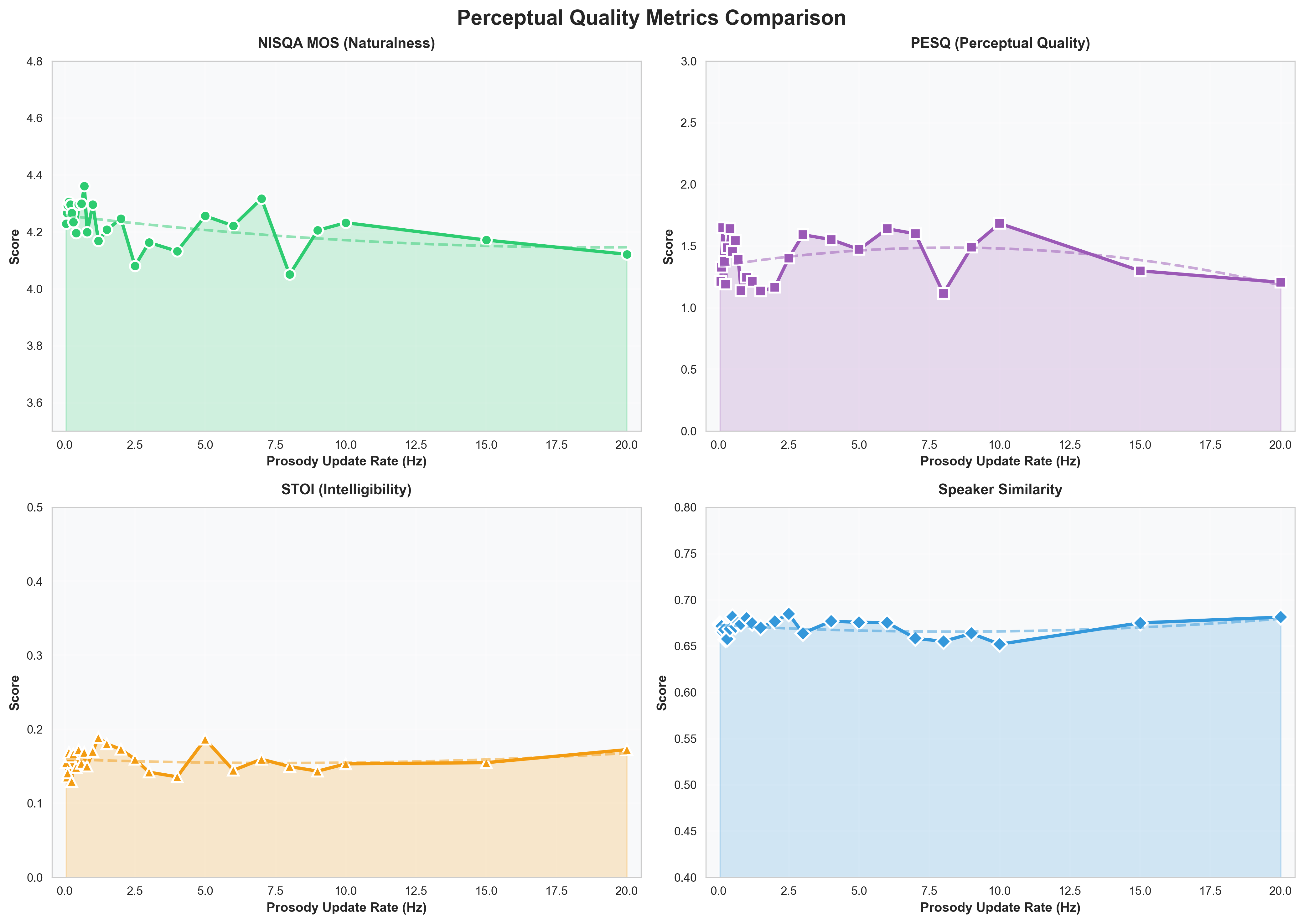}
\caption{Perceptual quality metrics (PESQ, STOI) demonstrating the bimodal phenomenon.}
\label{fig:prosody_c}
\end{subfigure}
\hfill
\begin{subfigure}[b]{0.48\textwidth}
\centering
\includegraphics[width=\textwidth]{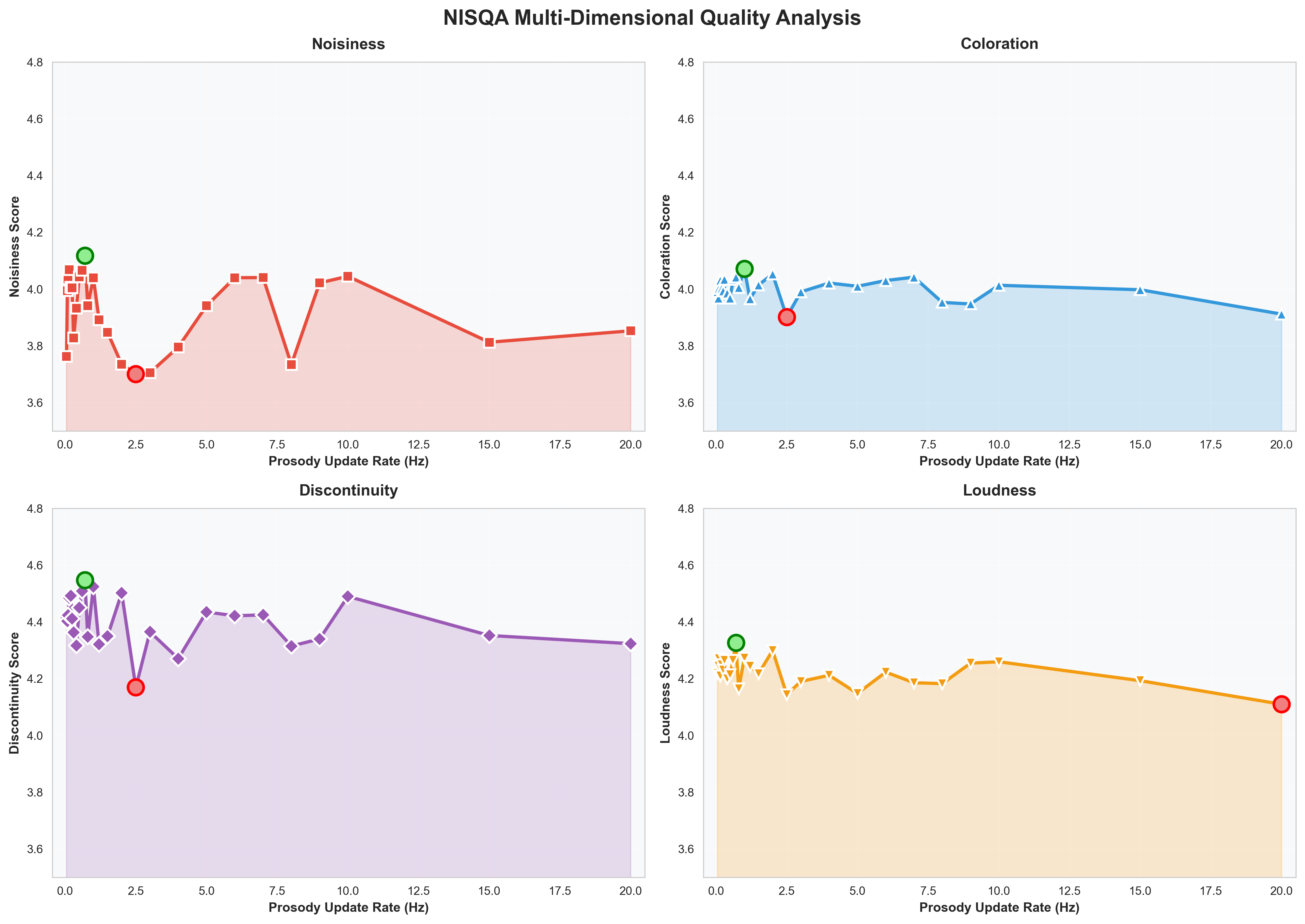}
\caption{NISQA dimensional breakdown revealing specific quality aspects affected by prosody sampling.}
\label{fig:prosody_d}
\end{subfigure}
\caption{Prosody sampling rate analysis across four key dimensions. The results reveal a bimodal quality distribution: high quality is achieved at both sparse rates (0.1--1 Hz) and dense rates ($>6$ Hz), while mid-range rates suffer from interpolation artifacts.}
\label{fig:prosody_analysis}
\end{figure*}

Figure~\ref{fig:prosody_analysis} presents the comprehensive results across four key dimensions. As shown in the top-left panel, bitrate increases approximately linearly with prosody sampling rate, rising from $\sim$312 bps at 0.05 Hz to $\sim$592 bps at 20 Hz (including amortized timbre). This linear relationship confirms that prosody transmission dominates the additional bandwidth consumption beyond the text baseline when prosody sampling rate is high.

The quality metrics exhibit a \textbf{bimodal distribution} with prosody sampling rate, as evident in the top-right and bottom-left panels. NISQA MOS scores show two distinct quality peaks separated by a substantial valley: a low-frequency peak clusters (0.05--1 Hz), a mid-frequency valley (1--5 Hz), and a high-frequency peak region (typically $>6$ Hz). We hypothesize this phenomenon arises from the interaction between prosody temporal granularity, TTS interpolation mechanisms, and perceptual salience of artifacts:

\textbf{Low-frequency (0.05--1 Hz):} Sparse updates align with sentence-level prosody. TTS interpolation generates smooth contours, and the model fills in natural micro-variations, achieving high quality with minimal bitrate.

\textbf{Mid-frequency (1--5 Hz):} This ``uncanny valley'' disrupts smooth interpolation with frequent but insufficient updates, causing perceptual discontinuities.

\textbf{High-frequency ($>6$ Hz):} Dense updates (e.g., 7--10 Hz) provide fine-grained control, overriding interpolation and bypassing discontinuity issues, but at significantly higher bitrate cost ($\sim$410 bps vs. 154 bps).

It is important to note that while the exact location of the low-frequency and high-frequency peaks may fluctuate, the overall bimodal trend remains consistent: both sparse and dense updates can yield high quality, whereas the mid-frequency range consistently degrades performance. While specific to our architecture, this suggests a fundamental trade-off in interpolation-based synthesis.

\textbf{Quality-bitrate considerations.} The low-frequency peaks achieve MOS $\sim$4.30--4.36 at 132--154 bps, offering exceptional efficiency. The high-frequency peak (observed around 7 Hz in our experiments) reaches MOS = 4.317 at 410 bps---comparable quality but at nearly $3\times$ the bandwidth cost. The mid-frequency valley is strictly dominated, with lower quality than both flanking peaks despite intermediate bitrate.

The PESQ and STOI metrics (bottom-left panel) largely corroborate the bimodal NISQA trend, though with greater variance. Speaker similarity remains stable across all rates, confirming that prosody sampling frequency affects \textit{expressiveness} rather than speaker identity.

Based on this bimodal distribution, we recommend the 0--1 Hz range as the universal operating point for all deployment scenarios. Although the high-frequency regime (e.g., $\sim$7 Hz) yields high absolute quality, the marginal perceptual improvement over the low-frequency peak is minimal and does not justify the significant bitrate penalty. Moreover, the high-frequency peak's location is volatile, varying with specific STT/TTS components and usage contexts, which complicates system tuning. By contrast, the 0--1 Hz range consistently maximizes the quality-to-bitrate ratio and ensures stability, whereas the mid-frequency range represents a design ``dead zone'' offering neither efficiency nor quality.

\textbf{Configuration Design Based on Analysis.} The insights from this prosody sampling rate analysis directly informed the design of our three operational modes. For the \textbf{minimal mode}, we selected 0.1 Hz from the low-frequency peak region, achieving maximum bandwidth efficiency while maintaining acceptable quality. This configuration targets extreme bandwidth constraints where every bit counts. For the \textbf{balanced mode}, we chose 0.5 Hz as a compromise that remains within the low-frequency efficiency region while providing more frequent prosody updates for improved expressiveness. This rate sits in between the low-frequency comfort zone, maximizing quality-per-bit while avoiding the discontinuity artifacts observed at 1--5 Hz. For the \textbf{high-quality mode}, we selected 1.0 Hz, which represents the upper boundary of the low-frequency regime. These carefully calibrated sampling rates explain why our three modes exhibit such narrow total bitrate variation as will be demonstrated in \ref{sec:benchmark}: prosody transmission contributes minimally even in high-quality mode, while compressed text dominates the bandwidth budget. This design philosophy prioritizes \textit{bandwidth efficiency} by leveraging TTS interpolation capabilities rather than brute-force transmission of dense prosody features.

\subsection{Benchmark Results}
\label{sec:benchmark}

Tables~\ref{tab:bitrate_comparison} and~\ref{tab:quality_comparison} present the comprehensive benchmark results comparing our three quality modes (minimal, balanced, and high-quality) against baseline codecs (Opus and EnCodec) and the Vevo framework. Table~\ref{tab:bitrate_comparison} shows the bitrate breakdown across different components, while Table~\ref{tab:quality_comparison} presents perceptual quality metrics. The high-quality mode is additionally evaluated under various noise conditions (0.1\%, 1\%, and 10\% bit error rates) to assess robustness to channel degradation.

\begin{table*}[t]
\centering
\caption{Bitrate comparison on LibriSpeech test-clean dataset. Values shown as mean $\pm$ standard deviation. Our system achieves $40\times$ to $50\times$ bitrate reduction compared to Opus (6 kbps) and $8\times$ reduction compared to EnCodec (1 kbps). Vevo* results from \cite{collette2024} with Zonos TTS and timbre update interval $d_{\text{timbre}} = 4$s, evaluated on a different test dataset and provided for reference only.}
\label{tab:bitrate_comparison}
\resizebox{0.7\textwidth}{!}{
\begin{tabular}{lrrrr}
\hline
\textbf{Method} & \textbf{Total (bps)} & \textbf{Text (bps)} & \textbf{Prosody (bps)} & \textbf{Timbre (bps)} \\
\hline
\multicolumn{5}{l}{\textit{Our System}} \\
minimal & 71.6$\pm$8.8 & 70.9$\pm$8.8 & 0.7$\pm$0.1 & 125.9$\pm$19.6 \\
balanced & 76.5$\pm$8.8 & 71.0$\pm$8.8 & 5.5$\pm$0.2 & 125.9$\pm$19.6 \\
high-quality & 79.6$\pm$8.9 & 65.8$\pm$8.9 & 13.9$\pm$0.2 & 251.8$\pm$39.2 \\
\hline
\hline
\multicolumn{5}{l}{\textit{Baseline Systems}} \\
Opus (6 kbps) & 6407.3$\pm$217.2 & --- & --- & --- \\
EnCodec (1 kbps) & 999.9$\pm$0.1 & --- & --- & --- \\
Vevo* & 650 & --- & --- & --- \\
\hline
\end{tabular}
}
\end{table*}

\begin{table*}[t]
\centering
\caption{Quality metrics comparison on LibriSpeech test-clean dataset. Values shown as mean $\pm$ standard deviation. Our system maintains comparable or superior perceptual quality (NISQA MOS) to the Vevo framework while achieving lower WER than EnCodec. Vevo* results from \cite{collette2024} with Zonos TTS and timbre update interval $d_{\text{timbre}} = 4$s, evaluated on a different test dataset and provided for reference only.}
\label{tab:quality_comparison}
\resizebox{0.95\textwidth}{!}{
\begin{tabular}{llrrrrr}
\hline
\textbf{Method} & \textbf{Noise Condition} & \textbf{WER} & \textbf{Speaker Sim} & \textbf{PESQ} & \textbf{NISQA MOS} & \textbf{STOI} \\
\hline
\multicolumn{7}{l}{\textit{Our System}} \\
minimal & --- & 0.259$\pm$0.204 & 0.673$\pm$0.085 & 1.238$\pm$0.381 & 4.280$\pm$0.348 & 0.150$\pm$0.036 \\
balanced & --- & 0.264$\pm$0.203 & 0.672$\pm$0.095 & 1.138$\pm$0.130 & 4.258$\pm$0.393 & 0.155$\pm$0.041 \\
high-quality & --- & 0.235$\pm$0.193 & 0.667$\pm$0.089 & 1.324$\pm$0.417 & 4.255$\pm$0.407 & 0.152$\pm$0.038 \\
\hline
\multicolumn{7}{l}{\textit{Noise Resilience (high-quality mode)}} \\
high-quality & No noise & 0.213$\pm$0.167 & 0.666$\pm$0.092 & 1.332$\pm$0.407 & 4.298$\pm$0.400 & 0.162$\pm$0.034 \\
high-quality & 0.1\% BER & 0.258$\pm$0.228 & 0.669$\pm$0.095 & 1.250$\pm$0.388 & 4.263$\pm$0.379 & 0.156$\pm$0.030 \\
high-quality & 1\% BER & 0.252$\pm$0.189 & 0.658$\pm$0.091 & 1.279$\pm$0.386 & 4.246$\pm$0.345 & 0.156$\pm$0.038 \\
high-quality & 10\% BER & 0.247$\pm$0.195 & 0.663$\pm$0.096 & 1.272$\pm$0.494 & 4.232$\pm$0.432 & 0.148$\pm$0.031 \\
\hline
\multicolumn{7}{l}{\textit{Baseline Systems}} \\
Opus (6 kbps) & --- & 0.032$\pm$0.048 & 0.673$\pm$0.058 & 2.284$\pm$0.349 & 2.455$\pm$0.371 & 0.906$\pm$0.027 \\
EnCodec (1 kbps) & --- & 0.110$\pm$0.092 & 0.450$\pm$0.074 & 1.334$\pm$0.112 & 2.083$\pm$0.393 & 0.805$\pm$0.029 \\
Vevo* & --- & 0.15 & 0.62 & 1.15 & 0.70 & 4.21 \\
\hline
\end{tabular}
}
\end{table*}

\subsubsection{Performance Evaluation}

\textbf{Bitrate Efficiency.} As shown in Table~\ref{tab:bitrate_comparison}, our system achieves remarkably low bitrates across all three quality modes. The total bitrate (excluding timbre amortization) ranges from 71.6 bps (minimal) to 79.6 bps (high-quality), representing a $40\times$ to $50\times$ reduction compared to Opus at 6 kbps and an $8\times$ reduction compared to EnCodec at 1 kbps. Notably, our bitrate is comparable to the Vevo framework ($\sim$650 bps) while maintaining similar perceptual quality. The timbre column shows the amortized bitrate for speaker embeddings over the test utterances (mean duration $\sim$12 seconds); however, in practical long-duration conversations, this cost amortizes to near-zero as discussed in Section~\ref{sec:prosody_timbre}. Furthermore, our timbre profile caching mechanism (Section~\ref{sec:prosody_timbre}) enables reuse of speaker embeddings across sessions, effectively eliminating timbre transmission overhead for familiar speakers in multi-party or recurring conversations. Therefore, the Total (bps) column represents the sustained bandwidth consumption for voice content transmission.

Our three quality modes exhibit minimal bitrate variation (71.6--79.6 bps), with differences primarily in prosody transmission frequency (0.7--13.9 bps) rather than text compression: text compression dominates the total bandwidth under relatively low prosody sampling rate (Section~\ref{sec:prosody_analysis}), while prosody contributes only marginally even in high-quality mode. Consequently, users can adopt the high-quality mode with negligible additional bandwidth cost ($<$10 bps overhead) while gaining substantial improvements in transcription accuracy (WER reduction from 0.259 to 0.235) and prosody fidelity. We therefore recommend the high-quality mode as the default configuration for most deployment scenarios.

\textbf{Perceptual Quality Metrics.} Table~\ref{tab:quality_comparison} presents the quality assessment results. Our system achieves speaker similarity scores (0.667--0.673) comparable to Opus (0.673) and substantially higher than EnCodec (0.450), indicating successful preservation of speaker identity through our embedding-based voice cloning approach. The PESQ scores (1.138--1.324) are competitive with EnCodec (1.334), though lower than Opus (2.284). This is expected, as PESQ is designed for waveform-level distortions in traditional codecs and does not fully capture the quality of synthesized speech. More importantly, our NISQA MOS scores (4.255--4.280) significantly exceed both Opus (2.455) and EnCodec (2.083), and match the Vevo framework (4.21). NISQA, as a non-intrusive deep learning-based metric trained on diverse speech quality dimensions, better reflects the perceptual naturalness of TTS-synthesized speech. The high NISQA scores confirm that our semantic compression approach produces highly natural and intelligible speech despite the ultra-low bitrate.

\textbf{STOI Analysis.} Our STOI scores (0.150--0.162) are notably lower than both Opus (0.906) and EnCodec (0.805), which might initially suggest poor intelligibility. However, this requires careful interpretation. STOI measures short-time objective intelligibility by computing temporal-spectral correlation between the original and reconstructed signals at the frame level (typically 10--30ms frames). It implicitly assumes \textit{frame-level temporal alignment} between reference and degraded audio---an assumption that holds for waveform codecs but fails for our TTS-based reconstruction approach.

In our system, the TTS synthesis process introduces \textit{temporal desynchronization} at multiple stages. First, the STT module may produce slightly different word boundaries than the original speech due to recognition uncertainty. Second, the TTS model generates speech with its own learned timing patterns conditioned on text and sparse prosody keyframes, which may not precisely match the original speaker's micro-timing (e.g., pause durations, consonant lengths, syllable boundaries). Third, our prosody features are transmitted at extremely low rates (0.1--1 Hz), providing only coarse temporal guidance rather than frame-by-frame alignment. As a result, even when the synthesized speech is highly intelligible and natural-sounding (as confirmed by high NISQA scores), the frame-by-frame waveform correlation measured by STOI remains low due to temporal shifts.

This phenomenon is inherent to semantic compression approaches that decouple content from acoustic realization. We include STOI in our evaluation not as a primary quality indicator, but to characterize the \textit{temporal alignment properties} of our reconstruction method and distinguish it from waveform-preserving codecs. For assessing actual intelligibility in semantic compression systems, metrics like WER (which measures content preservation) and NISQA (which evaluates perceptual naturalness) are more appropriate than STOI.

\subsubsection{Noise Resilience}

To assess the robustness of our system under degraded channel conditions, we evaluate the high-quality mode on the LibriSpeech test-clean dataset with simulated bit error rates (BER) of 0.1\%, 1\%, and 10\%, since it's been demonstrated that high-quality mode is fit to be the default configuration for most scenarios.

\textbf{Transcription Accuracy Degradation.} WER exhibits relative sensitivity to channel noise, rising from 0.213 (no noise) to 0.258 (0.1\% BER) by $\sim$20\%, then stabilizing around 0.247--0.252 at higher error rates. Interestingly, WER does not continue to degrade linearly at higher BERs (1\% and 10\%), and even shows slight improvement. We hypothesize this occurs because the receiver may rely more heavily on linguistic context and prosody cues to infer missing words, partially compensating for corrupted text data.

\textbf{Speaker Identity Preservation.} Speaker similarity remains remarkably stable across noise conditions, varying only slightly from 0.666 (no noise) to 0.658--0.669 under noise. This robustness arises from the amortized transmission strategy for speaker embeddings (Section~\ref{sec:prosody_timbre}). The 384-byte TIMBRE packet is transmitted only once at call initialization and re-sent only upon speaker change detection. This infrequent transmission makes timbre less susceptible to channel noise compared to continuously streamed data. Furthermore, speaker embeddings are transmitted with high priority and retransmission guarantees (Section~\ref{sec:transport}), ensuring reliable delivery even under noisy conditions. The minimal variation in speaker similarity confirms that voice identity is well-preserved regardless of channel quality.

\textbf{Perceptual Quality Preservation.} PESQ scores show slight degradation from 1.332 (no noise) to 1.250--1.279 under noise, a decline of $\sim$4--6\%. STOI decreases slightly from 0.162 to 0.148--0.156, and NISQA MOS exhibits a gradual downward trend from 4.298 to 4.232 as BER increases. These modest degradations (NISQA drops only $\sim$1.5\% even at 10\% BER) indicate that perceptual quality remains acceptable under noisy conditions. The graceful degradation can be attributed to our prioritized transmission strategy: TEXT packets receive high priority with retransmission, ensuring semantic content integrity, while low-priority prosody packets may be dropped under congestion. When prosody packets are lost, the receiver interpolates missing frames from adjacent keyframes (Section~\ref{sec:transport}), maintaining naturalness at the cost of reduced expressiveness. This design philosophy prioritizes \textit{intelligibility over expressiveness}---users can still understand the speech content even when fine-grained prosody is compromised. Even at 10\% BER---a severely degraded channel---the system maintains NISQA MOS above 4.2, indicating excellent perceptual quality and confirming the robustness of our semantic compression approach to channel impairments.

Our analysis reveals a divergence between semantic fidelity and perceptual quality: while transcription accuracy (WER) is sensitive to noise, perceptual quality (NISQA) remains remarkably resilient. This decoupling ensures that the reconstructed speech remains natural and clear even when semantic errors occur, a distinct advantage over traditional codecs where noise manifests as audible acoustic degradation.

\subsection{Computational Efficiency}

We evaluate the computational efficiency of our system by measuring the Real-Time Factor (RTF), defined as the ratio of processing time to audio duration. An RTF less than 1.0 indicates that the system processes audio faster than real-time, a critical requirement for live communication. Table~\ref{tab:rtf_analysis} presents the RTF results across different configurations and noise conditions.

\begin{table}[t]
\centering
\caption{Real-Time Factor (RTF) analysis on LibriSpeech test-clean dataset. Experiments conducted on a single NVIDIA RTX 4080 GPU with Intel Core i9-13900K CPU. RTF $<$ 1.0 indicates faster than real-time processing.}
\label{tab:rtf_analysis}
\resizebox{0.85\columnwidth}{!}{
\begin{tabular}{llc}
\hline
\textbf{Config} & \textbf{Noise} & \textbf{RTF} \\
\hline
minimal\_mode & No noise & 0.396$\pm$0.080 \\
balanced\_mode & No noise & 0.404$\pm$0.074 \\
high\_quality\_mode & No noise & 0.387$\pm$0.046 \\
\hline
\end{tabular}
}
\end{table}

Our system consistently achieves an RTF of approximately 0.4 across all modes, meaning it requires only 40\% of the audio duration to process the full pipeline (STT, compression, transmission, decompression, and TTS). This performance demonstrates that the system is well-suited for real-time deployment, leaving ample headroom for other concurrent tasks.

One of our core design choices is that \textit{we explicitly trade computational power for bandwidth efficiency.} By leveraging GPU acceleration for the neural components (STT and TTS), we achieve ultra-low bitrates ($\sim$80 bps) that would be impossible with traditional lightweight codecs. This design choice reflects the economic reality of our target scenarios (e.g., maritime, satellite), where bandwidth is the scarce and expensive resource, while computational power (even if requiring a GPU) is mostly a one-time capital investment that is relatively inexpensive compared to the recurring operational cost of satellite data.

\subsection{End-to-End Latency Analysis}

While Real-Time Factor (RTF) measures throughput, \textit{latency} determines the delay between a speaker uttering a word and the listener hearing it. Our system's latency consists of three primary components:

\textbf{1. Sender-Side Algorithmic \& Processing Latency.} The STT module operates on 400ms audio chunks with a 50ms overlap. This imposes a minimum algorithmic delay of 400ms (waiting for the chunk to fill). Processing this chunk takes approximately $400\text{ms} \times \text{RTF}_{\text{STT}}$. With an STT RTF of $\sim$0.15 (on GPU), this adds $\sim$60ms. VAD buffering adds a dynamic component, typically waiting 250ms to confirm speech onset. Thus, the total sender-side latency is approximately $400 + 60 = 460$ms.

\textbf{2. Transmission Latency.} This is network-dependent. In satellite scenarios, propagation delay can range from 250ms (GEO) to 40ms (LEO). Our ultra-low bitrate ($\sim$80 bps) ensures that serialization delay is negligible, and packets are small enough to avoid fragmentation delays.

\textbf{3. Receiver-Side Synthesis Latency.} The TTS model (XTTS-v2) supports streaming synthesis. However, to ensure natural prosody, it typically requires a minimal context window (e.g., 3--5 words). Assuming a speaking rate of 3 words/sec, this adds a buffering delay of $\sim$1 second. The synthesis process itself, with an RTF of $\sim$0.2, adds minimal additional delay for the first audio chunk (Time to First Audio).

Combining these factors, the theoretical minimum end-to-end latency is approximately $1.5$--$2.0$ seconds (excluding network propagation). While higher than standard telephony ($\sim$200ms), this is a necessary trade-off for achieving 80 bps communication. In half-duplex "push-to-talk" modes (common in tactical/maritime radio), this latency is masked by the turn-taking protocol and is perceptually acceptable.

\subsection{Discussion}

Our experimental evaluation demonstrates that the STCTS pipeline achieves ultra-low bitrate voice communication ($\sim$80 bps sustained bandwidth) while maintaining high perceptual quality comparable to state-of-the-art semantic compression systems. It is noteworthy that STCTS represents an extreme audio compression approach, achieving compression ratios exceeding 3000:1 compared to uncompressed PCM audio (16 kHz, 16-bit: 256 kbps $\rightarrow$ 80 bps), and 80:1 even against highly optimized traditional codecs like Opus at 6 kbps. This radical compression is enabled by discarding acoustic waveform fidelity entirely and reconstructing speech from purely semantic and prosodic representations, fundamentally redefining the trade-off between bitrate and perceptual quality in voice communication. Beyond bitrate efficiency and perceptual quality, our explicit semantic decomposition approach offers several architectural advantages over end-to-end neural codecs and token-based semantic compression.

\textbf{Compute-Bandwidth Trade-off.} A core design principle of STCTS is the strategic exchange of computational power for bandwidth efficiency. While traditional codecs minimize compute to run on minimal hardware, we leverage modern accelerators (e.g., GPUs, NPUs) to perform sophisticated semantic analysis and synthesis, thereby reducing bandwidth consumption by orders of magnitude. This trade-off is economically advantageous in our target scenarios (maritime, satellite, tactical), where bandwidth is the scarce, recurring cost (e.g., \$10/MB), whereas computational hardware is a one-time fixed cost. Our evaluation confirms that with a single consumer GPU (RTX 4080), the system runs comfortably faster than real-time (RTF $\sim$0.4), validating the feasibility of this approach.

\textbf{Privacy-Preserving End-to-End Encryption.} The textual intermediate representation enables straightforward application of standard encryption protocols (e.g., AES-256, RSA) to protect semantic content during transmission. Since text, prosody features, and speaker embeddings are explicitly structured data, they can be encrypted independently with different keys or access policies, enabling fine-grained privacy controls. For instance, text content can be encrypted end-to-end between callers, while prosody features (which convey emotion but not semantic content) might use a weaker encryption level or remain unencrypted for network optimization. This flexibility contrasts with neural codec latent representations, which are entangled high-dimensional vectors that resist selective encryption. Furthermore, the explicit text representation facilitates compliance with data protection regulations (e.g., GDPR right to explanation), as transmitted content is human-interpretable rather than opaque neural activations.

\textbf{Modular Design and Model Upgrade-ability.} The decoupled STCTS pipeline allows independent upgrading of each component without retraining the entire system. As more accurate STT models emerge (e.g., future Whisper versions, domain-specific ASR), they can be seamlessly integrated by replacing the STT module while retaining existing compression and TTS components. Similarly, advances in TTS (e.g., improved voice cloning, lower-latency synthesis) can be adopted without modifying the upstream pipeline. This modularity also enables domain-specific optimization: medical consultation systems can employ specialized medical STT models and terminology-aware text compression, while casual conversation systems use general-purpose models. In contrast, end-to-end neural codecs require full model retraining to incorporate improvements, and their monolithic architecture resists task-specific customization.

\textbf{Inherent Interpretable Intermediate Representation.} The explicit textual representation provides transparency and debuggability absent in neural codec latent spaces. System developers can inspect transmitted text to diagnose transcription errors, measure content-level bitrate allocation, and implement content-aware optimizations (e.g., domain-specific dictionaries, phrase prediction). Users can optionally view transcriptions in real-time for accessibility (e.g., hearing-impaired communication) or quality assurance (e.g., verifying critical instructions in aviation or telemedicine). This interpretability also enables secondary applications: conversation logging, sentiment analysis, automatic summarization, and multilingual translation—all operating on the transmitted text stream without additional processing. Token-based semantic codecs (e.g., Vevo's discrete audio tokens) lack this human-interpretable intermediate form, limiting their utility beyond speech reconstruction.

\textbf{Limitations.} Despite its advantages, the STCTS approach has inherent limitations. First, we evaluate STCTS on LibriSpeech, a corpus of read audiobooks. While this provides a standardized benchmark for reconstruction quality, it does not capture the complex dynamics of spontaneous conversational speech (e.g., turn-taking, interruptions, overlapping speech, disfluencies). In real-world deployments, these conversational phenomena might pose challenges for the VAD and STT modules that are not reflected in our current benchmarks. Second, the system incurs a relatively high end-to-end latency ($\sim$1.5--2.0 seconds) compared to traditional waveform codecs. This latency, inherent to the semantic processing chain (STT buffering, text generation, TTS synthesis), makes the system less suitable for rapid-fire, full-duplex interruptions, although it remains acceptable for half-duplex or high-latency scenarios (e.g., satellite PTT). Third, the system is designed strictly for speech; non-speech acoustic events (e.g., laughter, crying, background music) are filtered out or ignored, resulting in their loss at the receiver. Fourth, reconstruction quality is bounded by STT/TTS performance; transcription errors can lead to semantic deviations. Finally, performance depends on model availability for the target language.

\textbf{Directions for Future Improvement.} While our current system demonstrates strong performance, several architectural enhancements could further improve quality and efficiency. First, \textit{advanced text compression} using large language models (LLMs) could achieve near-optimal entropy coding. Recent work \cite{deletang2024,lmcompress} shows that probabilistic language models can drive arithmetic coding to compress text to $\sim$30\% of traditional compressor sizes. Integrating a lightweight on-device LM or leveraging cloud-based LLMs for predictive compression could reduce our text bitrate from $\sim$70 bps to $\sim$20--30 bps, bringing total bandwidth below 50 bps. Second, \textit{adaptive prosody transmission} could dynamically adjust update rates based on speech content: increasing frequency during emotionally expressive segments or rapid pitch changes, while reducing to near-zero during monotone speech. This content-aware strategy could maintain high expressiveness while further minimizing bandwidth. Third, \textit{joint optimization of STT and TTS models} through multi-task learning or knowledge distillation could improve end-to-end reconstruction quality. For instance, training the TTS model to predict not only speech but also STT-generated transcriptions could teach it to compensate for common transcription errors, reducing WER degradation. Finally, \textit{neural codec hybridization} could combine our semantic approach with residual waveform coding: transmit text and prosody semantically (as we do), but add a low-bitrate neural codec stream ($\sim$100--200 bps) to encode fine-grained acoustic details (e.g., breathing, laughter, background ambience) that semantic compression discards. This hybrid approach could improve STOI and PESQ scores while maintaining ultra-low total bitrate.

The architectural advantages and future improvement directions position STCTS as a versatile framework for diverse communication scenarios beyond the traditional telephony use case, including privacy-sensitive applications (e.g., secure government communication) and resource-constrained deployments (e.g., satellite IoT networks).

\section{Conclusion}

This paper presented STCTS, an ultra-low bitrate voice communication system achieving $\sim$80 bps—a $75\times$ reduction over traditional codecs—while maintaining high perceptual quality (NISQA MOS $>$4.2) through the orthogonal decomposition of speech into linguistic content, prosodic expression, and speaker identity. Our evaluation demonstrates that explicit semantic modeling enables robust communication in bandwidth-constrained environments, with key findings highlighting the effectiveness of sparse prosody interpolation ($<$14 bps) and the system's resilience to channel noise. While the approach introduces higher latency ($\sim$1.5--2.0s) and relies on generative reconstruction, it offers significant advantages in modularity, privacy-preserving encryption, and interpretability compared to end-to-end neural codecs. Future work will focus on integrating LLM-based text compression and adaptive prosody transmission to further optimize the trade-off between bitrate, latency, and naturalness.

\appendices
\onecolumn
\section{Quality Mode Specifications}
\label{app:quality_modes}

The predefined quality modes are:

\textbf{Minimal Mode.} Employs small STT model* with aggressive Brotli compression (level 9*) and text preprocessing*. Prosody updates at 0.1 Hz* transmitting only pitch feature* with minimal quantization (3-bit pitch*, 2-bit energy*, rate disabled*). Uses 192-dimensional speaker embedding* at float16 precision* with change detection threshold 0.4*. Designed for extreme bandwidth constraints such as legacy satellite links, 2G networks, or congested mobile connections where intelligibility takes priority over naturalness.

\textbf{Balanced Mode.} Uses small STT model* with Brotli compression (level 5*) balancing compression speed and efficiency. Prosody updates at 0.5 Hz* (every 2 seconds) with pitch, energy, and speaking rate features* quantized to 6-bit*, 5-bit*, and 5-bit* respectively. Includes emotion tracking at 0.2 Hz*. Uses 192-dimensional speaker embedding* (float16*) with change detection threshold 0.3*. This is the default mode providing the best quality-to-bitrate ratio, achieving substantially lower bitrates than neural audio codecs like Lyra \cite{lyra} or EnCodec \cite{encodec} while maintaining excellent perceptual quality.

\textbf{High Quality Mode.} Employs distil-large-v3 STT model* for maximum transcription accuracy with Brotli compression (level 5*). Prosody updates at 1.0 Hz* (every second) with all features* at high precision (8-bit pitch*, 6-bit energy*, 6-bit speaking rate*). Uses 192-dimensional speaker embedding* at full float32 precision* with stricter change detection threshold 0.25*. Disables text preprocessing* to preserve exact transcription. Additional audio processing parameters include 300ms chunk duration* and 0.4 VAD threshold*. Designed for stable 3G/4G/WiFi connections where accuracy and naturalness are prioritized over bandwidth efficiency.

Users can manually select the mode or enable adaptive switching based on measured network throughput and latency. The three-tier configuration provides clear tradeoffs: minimal mode maximizes compression for constrained networks, balanced mode optimizes the quality-bitrate ratio for typical scenarios, and high-quality mode prioritizes accuracy and naturalness when bandwidth permits. Custom configurations can be created by copying and modifying the YAML files, allowing fine-grained control over the bitrate-quality tradeoff for specific deployment scenarios.

\end{document}